# Geospatial AI for Liquefaction Hazard and Impact Forecasting: A Demonstrative Study in the U.S. Pacific Northwest

Morgan D. Sanger*[1] and Brett W. Maurer[2]

**Abstract**

Recent large-magnitude earthquakes have demonstrated the damaging consequences of soil liquefaction and reinforced the need to understand and plan for liquefaction hazards at a regional scale. In the United States, the Pacific Northwest is uniquely vulnerable to such consequences given the potential for crustal, intraslab, and subduction zone earthquakes. In this study, the liquefaction hazard is predicted geospatially at high resolution and across regional scales for 85 scenario earthquakes in the states of Washington and Oregon. This is accomplished using an emergent geospatial model that is driven by machine learning, and which predicts the probability of damaging ground deformation by surrogating state-of-practice geotechnical models. The adopted model shows improved performance and has conceptual advantages over prior regional-scale modeling approaches in that predictions (i) are informed by mechanics, (ii) employ more geospatial information using machine learning, and (iii) are geostatistically anchored to known subsurface conditions. The utility of the resulting predictions for the 85 scenarios is then demonstrated via asset and network infrastructure vulnerability assessments. The liquefaction hazard forecasts are published in a GIS-ready, public repository and are suitable for disaster simulations, evacuation route planning, network vulnerability analysis, land-use planning, insurance loss modeling, hazard communication, public investment prioritization, and other regional-scale applications.

**Keywords**

Geospatial machine learning; regional liquefaction hazard mapping; impact forecasts.

[1] Department of Civil & Environmental Engineering University of Washington, 3760 E. Stevens Way NE Seattle, WA 98195, sangermd@uw.edu, *Corresponding author

[2] Department of Civil & Environmental Engineering University of Washington, 3760 E. Stevens Way NE Seattle, WA 98195, bwmaurer@uw.edu

## Introduction

Earthquake-induced consequences in terms of loss of life and property, infrastructure damage, and service disruptions realized in recent events have underscored the need to forecast and plan for these hazards on a regional scale. For example, the 2023 Kahramanmaras, Turkey sequence caused more than 55,000 deaths in Turkey and Syria, and over 37,000 building collapses in Turkey (British Red Cross, 2024). In New Zealand, the 2010-2011 Canterbury sequence saw more than 185 deaths and realized approximately NZ$30 billion in total insured losses, largely due to liquefaction-related damages (Ministry for Culture and Heritage, 2023; Parker and Steenkamp, 2012). Enumerable seismically active regions across the world are vulnerable to similar earthquake hazards and losses. The Pacific Northwest region of the United States – particularly coastal Washington (WA) and Oregon (OR) – presents one such region with the potential for crustal, intraslab, and subduction zone earthquakes. A preliminary assessment by the United States Department of Homeland Security estimated an average annualized earthquake loss of $1.94 billion in WA and OR (Jaiswal et al., 2023). However, this estimate neglects damages due to liquefaction- and coseismic landslide-related ground failure. As demonstrated in Turkey and New Zealand, well-characterized liquefaction hazard at a regional scale is needed for damage forecasting, earthquake preparedness, and response. Making regional scale predictions with state-of-practice liquefaction models is currently infeasible, however, given these models are conditioned on in-situ geotechnical data (e.g., cone penetration testing, "CPT," measurements) and cannot be deployed continuously over large areas. Toward this end, "geospatial" models for predicting liquefaction have emerged (e.g., Azul et al., 2024; Bozzoni et al., 2021; Bullock et al., 2023; Jena et al., 2023; Kim, 2023; Rashidian and Baise, 2020; Todorovic and Silva, 2022; Zhu et al., 2017).

Geospatial liquefaction models are generally conditioned on geospatial proxies for liquefaction hazard derived from existing maps, models, and remote-sensing datasets (e.g., geology maps, depth to groundwater models, topographic variables derived from aerial measurements). Recent tests demonstrate promising potential (e.g., Geyin et al., 2020) and expanded utility offered by geospatial liquefaction models, such as informing reconnaissance efforts immediately after an earthquake (e.g., Asadi et al., 2024). The United

States Geological Survey (USGS) has recognized this potential and adopted the seminal model of Zhu et al. (2017), later refined by Rashidian and Baise (2020), in their post-earthquake information products (Allstadt et al., 2022). However, predicting liquefaction without site-specific subsurface data remains challenging, and current geospatial models exhibit practical and conceptual limitations (Geyin et al., 2022). Namely, they tend to: (i) directly predict liquefaction manifestation as a binary or probability of manifestation at the ground surface without consideration of established liquefaction mechanics; (ii) exclude available subsurface geotechnical data; (iii) use relatively little of the publicly available geospatial data; (iv) are trained via traditional statistical methods (e.g., logistic regression) which cannot capture complex, nonlinear relationships; and (v) are not provided as a packaged model (e.g., as a code repository or executable software) (Sanger et al., 2025).

Accordingly, Sanger et al. (2025) developed a novel approach to geospatial liquefaction modeling designed to address each of these limitations. In this approach, geospatial "surrogate" models were trained to mimic geotechnical models at CPT measurement sites using machine learning (ML) and a large library of geospatial predictor variables. In this way, the geospatial models benefit from the knowledge of liquefaction mechanics embedded in the geotechnical models they surrogate. The ML model predictions are geostatistically updated by subsurface data, such that liquefaction predictions proximal to in-situ geotechnical data are driven by measured subsurface conditions and geotechnical models, and distal predictions are informed by the ML surrogate models. The models are packaged as mapped parameters, which – when applied with event-specific ground-motion information using the provided code repository – rapidly produce probabilistic predictions of liquefaction-induced ground failure across the affected region. Thus, in contrast to the common limitations of prior geospatial models, the Sanger et al. (2025) approach was informed by mechanistic knowledge, used geotechnical data both in model training and local updating, employed a large volume of geospatial data, aimed to exploit that information via ML, and is very simple for users to apply in the forward direction, both to future scenario events and in near-real-time.

In this study, we demonstrate the utility of the Sanger et al. (2025) geospatial liquefaction models (hereafter "GLMs") for hazard and impact forecasting using the U.S. Pacific Northwest as a study region.

*First*, we test the GLMs in the region against the existing state-of-practice geospatial model using case histories from the 28 February 2001 M6.8 Nisqually, WA, earthquake. *Next*, to deploy the GLMs for regional-scale hazard assessment, we must consider that the seismic setting of the Pacific Northwest is one of infrequent but potentially moderate-to-large magnitude earthquakes, where few significant events have occurred within the historic observation period. Such regions are particularly reliant on scenario earthquake simulations for disaster planning, land-use regulation, loss modeling, network vulnerability analysis, hazard communication, etc. To this end, we implement the GLMs for 85 scenario earthquakes in the region and produce high-quality digital maps. *Then*, we consider the modeled liquefaction hazard and impacts through event-specific and holistic contexts (i.e., all available scenarios) to illustrate the utility of the GLMs in various region-scale contexts.

## Data and Modeling

### *Geospatial Liquefaction Modeling*

As previously introduced, the Sanger et al. (2025) modeling framework fundamentally differs from prior geospatial liquefaction models. A total of 37 geospatial variables (i.e., features) were used to represent soil thickness, saturation, density, and/or typology. These features include measurements derived from elevation models and hydrologic datasets (e.g., topographic slope, distance to rivers of different flow-orders), and predictions made by other models (e.g., sand fraction, water content). The GLMs were developed to be globally applicable, requiring that the selected geospatial variables are available on a global scale. This precludes data that is only available on regional or national scales, such as state geologic maps, which may have higher resolution or more regional specificity. Although regionally specific models may benefit from less variance between variables and targets, global models can train on more data. Testing by Sanger et al. (2025) compared a New Zealand-specific model and the global GLMs against observational data, and the two models did not demonstrate statistically significant performance differences.

As further discussed momentarily, the Sanger et al. (2025) GLMs trained and tested on more than 37,000 CPTs collectively sourced from publicly available regional, national, and global datasets (Rasanen et al. 2024; USGS, 2019; Sanger et al. 2024a; Regione Emilia-Romagna, 2024; New Zealand EQC, 2016;

Geyin and Maurer, 2021a). The collection of geotechnical and geospatial data employed by this model is summarized in Table 1. In the Pacific Northwest, the model training data is largely concentrated in the western part of the region (e.g., Seattle, WA and Portland, OR) (Fig. 1). While the eastern Pacific Northwest is less represented in the training data feature space, the western part of the region – where population, infrastructure, and seismic hazard are concentrated – sees better representation and is anchored to known subsurface conditions.

Existing geospatial models typically predict liquefaction response using case history observations, or locations of documented liquefaction manifestation such as ejecta at the ground surface, following an earthquake. The Sanger et al. (2025) GLMs instead derive the expected liquefaction response from CPT measurements. Given the relative infrequence of large magnitude earthquakes capable of causing liquefaction, the size and growth potential of case history datasets is small compared to the size and growth potential of measured subsurface data, which are collected routinely and are increasingly available as published datasets (e.g., Ulmer et al, 2023). The GLMs were developed to surrogate three CPT-based state-of-practice geotechnical models: the liquefaction potential index (*LPI*) (Iwasaki et al., 1978); a modified *LPI*, termed $LPI_{ISH}$ (Maurer et al., 2015); and the liquefaction severity number (*LSN*) (Van Ballegooy et al., 2014). These are the most widely adopted CPT-based geotechnical manifestation models in geotechnical earthquake engineering research and practice, as described by Ulmer et al., 2024, among others. Other models have been proposed (e.g., Upadhyaya et al., 2022; Ulmer et al., 2024, Hu and Liu, 2019; Wang and Hu, 2025, among others), which are not yet widely adopted but will be considered in future research. Modeling distinctions and calculation details for the three geotechnical models are summarized in *Supplemental Information*. These models each output an index (often called a "vulnerability index", but here referred to as "manifestation index", *MI*) used to predict a soil profile's cumulative liquefaction response, or damage potential, at the ground surface. In this way, *LPI*, $LPI_{ISH}$, and *LSN* are pragmatic proxies of near-surface damage potential. Fragility functions conditioned on each of these indices can then be used to estimate the probabilities of certain outcomes, including ground failure (i.e., deformation and ejecta) (Geyin and Maurer, 2020), pipeline rupture (Toprak et al., 2019), foundation damage (Maurer et al., 2025), or other liquefaction consequences of interest. The final products (e.g., probability of ground failure maps) can be used by engineers, city planners, and other stakeholders to inform land-use planning, zoning, interrogation of evacuation and emergency response routes, identification of vulnerable infrastructure, and more, as demonstrated in *Liquefaction Hazard and Impact Forecasting for Scenario Earthquakes*. The surrogate modeling workflow is illustrated in Fig. 2 and is described here in brief to inform the application of the models in hazard and impact forecasting. Complete model development and performance details are provided by Sanger et al. (2025).

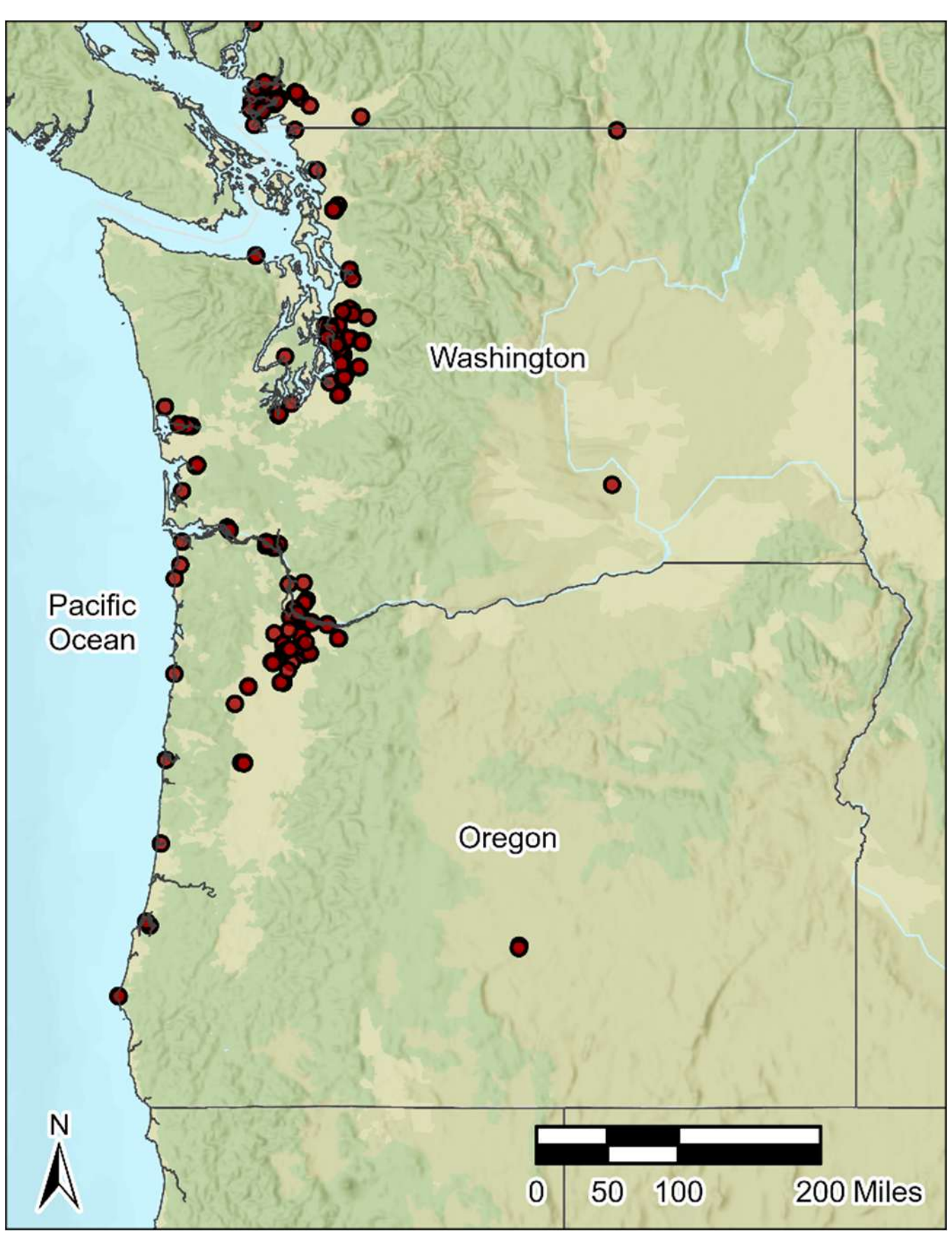

**Fig. 1.** Spatial distribution of CPT training and test data in the Pacific Northwest. (Sources: Esri and USGS.)

**Table 1.** Summary of employed geotechnical and geospatial data.

| Variable | | Source |
|---|---|---|
| *Targets* | | |
| CPTs | Global | Geyin and Maurer (2021a) |
| | Italy | Regione Emilia-Romagna (2024) |
| | New Zealand | New Zealand EQC (2016) |
| | North America | Sanger et al. (2024a) |
| | Pacific Northwest | Rasanen et al. (2024) |
| | United States | USGS (2019) |
| *Features* | | |
| Bulk density | | Hengl (2018a) |
| Clay fraction | | Hengl (2018b) |
| Compound topographic index | | Amatulli et al. (2020) |
| Convergence | | Amatulli et al. (2020) |
| Depth to bedrock | | Shangguan et al. (2017) |
| Depth to groundwater | | Fan et al. (2013) |
| Distance to coast | | NASA (2020) |
| River distances (Flow orders 1-2 to 1-8) | | Lehner and Grill (2013) |
| Elevation standard deviation | | Amatulli et al. (2020) |
| Geomorphon | | Amatulli et al. (2020) |
| Height above nearest drainage | | Nobre et al. (2011) |
| Landform entropy | | Amatulli et al. (2018) |
| Landform Shannon index | | Amatulli et al. (2018) |
| Landform uniformity | | Amatulli et al. (2018) |
| Major | | Amatulli et al. (2018) |
| Maximum multiscale deviation | | Amatulli et al. (2020) |
| Maximum multiscale roughness | | Amatulli et al. (2020) |
| Profile curvature | | Amatulli et al. (2020) |
| Roughness | | Amatulli et al. (2020) |
| Sand fraction | | Hengl (2018c) |
| Scale of the maximum multiscale deviation | | Amatulli et al. (2020) |
| Scale of the maximum multiscale roughness | | Amatulli et al. (2020) |
| Silt fraction | | Hengl (2018d) |
| Soil class | | Hengl and Nauman (2018) |
| Tangential curvature | | Amatulli et al. (2020) |
| Terrain ruggedness index | | Amatulli et al. (2020) |
| Topographic position index | | Amatulli et al. (2020) |
| Topographic slope | | Amatulli et al. (2020) |
| Vector ruggedness measure | | Amatulli et al. (2020) |
| Time-averaged shear wave velocity ($V_{s30}$) | | Heath et al. (2020) |
| Water content | | Hengl and Gupta (2019) |

To develop the GLMs, each of the compiled ~37,000 CPT profiles were subjected to peak ground accelerations (*PGA*s) of 0.05 g to 2.0 g and rupture magnitudes (i.e., moment magnitude, M) of 4.5 to 9.0, resulting in arrays of *MI* (i.e., *LPI*, $LPI_{ISH}$, and *LSN*; Eq. 1 to Eq. 3, Fig 2a) as a function of *PGA.* In this paper, *PGA* is always defined as the magnitude-scaled *PGA* ("$PGA_M$") according to the magnitude-scaling factor of Idriss and Boulanger (2008). Each profile and each computed $MI\text{-}PGA_M$ array were fit with unique fitting parameters *A* and *B* (Eq. 4, Fig 2a), which ultimately become the prediction targets of the ML model. Then, the library of geospatial variables is sampled at each CPT location to comprise the feature variables of the ML model (Fig. 2b). Discussion of the relative feature importance of the geospatial variables, their distributions, and domain constraints on predictions are also described in Sanger et al. (2025).

The relationships between the surrogate geotechnical models (i.e., *A* and *B*) and the geospatial features are exploited using supervised ML (Fig. 2c). During ML model development, care was taken to adhere to best modeling practices and to avoid the serious modeling flaws exhibited by many existing ML liquefaction models, as documented by Maurer and Sanger (2023). A variety of algorithms were iteratively evaluated according to the training, cross validation, and test-set performances and overfitting behavior, from which a bagged decision tree ensemble architecture was ultimately selected (Breiman, 1996; 1999). Using a 90%:10% data split, an ML model for each of the six targets (*A* and *B* for *LPI*, $LPI_{ISH}$, and *LSN*) was independently optimized using parallelized hyperparameter grid search to minimize a ten-fold cross-validation mean squared error (MSE) loss function. The final trained models were deployed globally at ~90 m resolution (~0.000833°), excluding areas outside the scope and domain of the model, such as topographic slopes >5° (Amatulli et al., 2020), lakes (Messager et al., 2016), glaciers (RGI Consortium, 2023), the Greenland Ice Sheet (Lewis, 2009), and permafrost regions (Brown et al., 2002) (Fig. 2d).

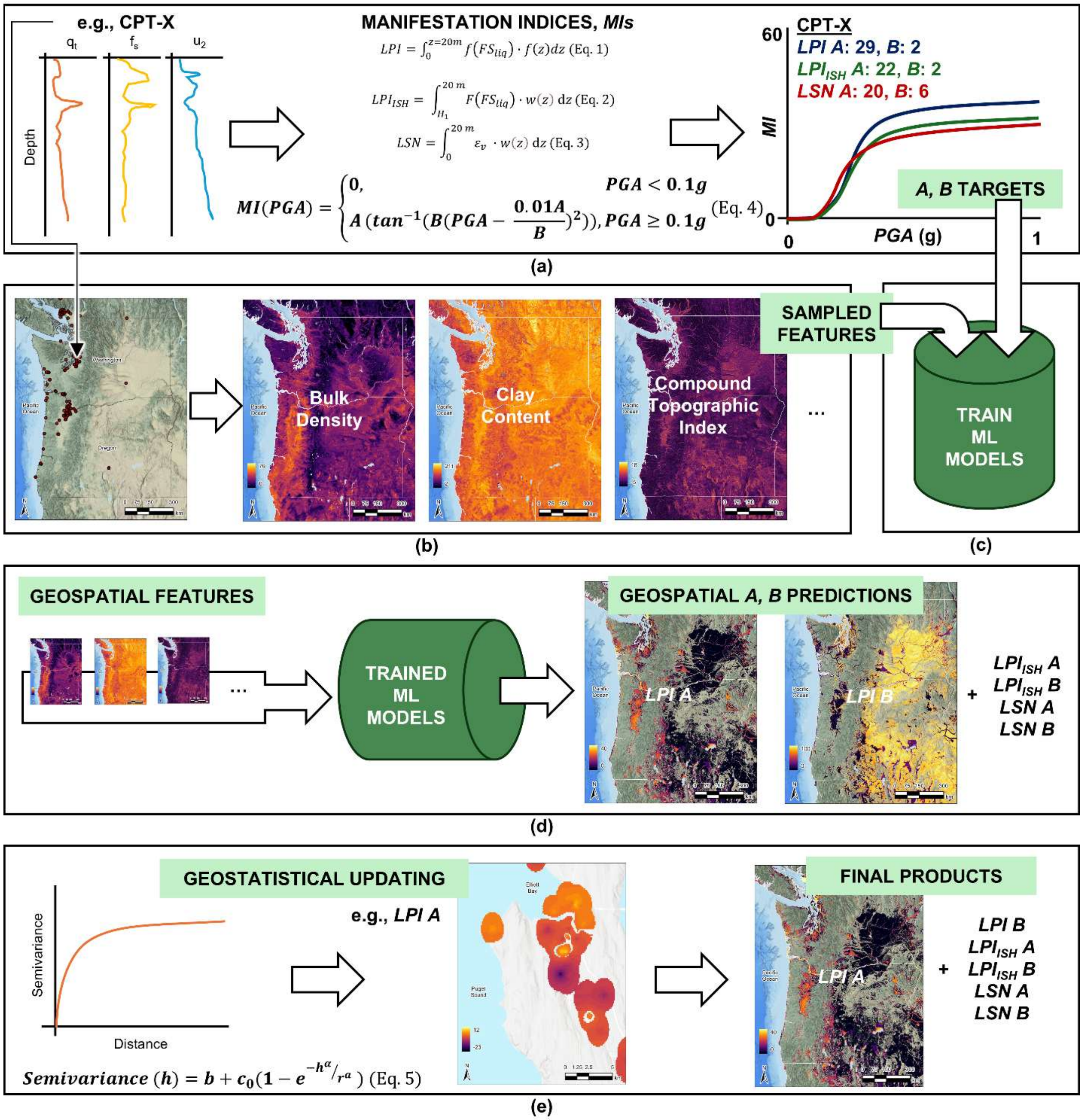

**Fig. 2.** Model development workflow for the GLMs, including (a) deriving ML model targets *A* and *B* for *LPI*, $LPI_{ISH}$, and *LSN* from compiled CPT profiles; (b) sampling ML model features from compiled geospatial variables; (c) training the supervised ML models; (d) deploying the trained ML models geospatially; and (e) geostatistically updating the ML predictions via regression kriging. Further descriptions and definitions of the equations (Eq. 1 through Eq. 5) are provided in the *Supplemental Information*.

ML predictions were then refined using regression kriging (e.g., Hengl et al., 2007), which locally corrects predictions in the vicinities of measured subsurface conditions using spatial interpolation of residuals (Fig. 2e). The spatial kriging function is described with a semivariogram (Eq. 5, Fig 2e). After geostatistical updating, final global predictions of *A* and *B* for each *MI* are provided as raster files by Sanger et al. (2024b). When combined with $PGA_M$, *A* and *B* produce event-specific predictions of *LPI*, $LPI_{ISH}$, and *LSN*, which can be implemented in any geospatial calculator (e.g., ArcMap, QGIS, Python, Matlab). A simple script is provided by Sanger et al. (2024c) to implement any of the developed models in Python (Jupyter Notebook) and Matlab within the DesignSafe computing environment (Rathje, 2017). In this way, the expected liquefaction response has been precomputed globally for all possible earthquakes and can be queried by the end user without high performance computing capabilities or ML expertise. The resulting event-specific predictions of *LPI*, $LPI_{ISH}$, and *LSN* can be propagated via fragility functions, or “damage” functions (e.g., Geyin and Maurer, 2020, Toprak et al., 2019, Maurer et al., 2025). The three geotechnical models have different conceptual foundations, as summarized in *Supplemental Information*, and thus they may differ in their efficacies; however, there are generally too few case-history observations to establish consensus on which model performs best (e.g., Geyin et al., 2020; Rasanen et al., 2023). For these reasons, users may wish to interpret the predictions independently or as the average of the predictions from one or more of the models. The performance of the GLMs in predicting case-history observations, both globally and in the Pacific Northwest, is described in *Results and Discussion*.

***Scenario Earthquakes in the Pacific Northwest***

The GLMs can be deployed either at discrete locations or across the extent of an affected area, and predictions can be made either in near-real-time after an event using recorded ground motions, or in planning for a scenario event using predicted motions. A scenario represents one realization of a potential earthquake by estimating ground motions for an assumed fault-rupture magnitude, location, and geometry. Scenarios are commonly used in planning, coordinating emergency response, and examining exposure of infrastructure at regional scale. The USGS has nearly 800 scenario ground-motion simulations for the continental U.S. In this study, we employ all available scenarios for finite-fault ruptures in the states of WA and OR to conduct regional hazard and impact forecasting. We simulate 85 scenarios in total, summarized in Table 2, from three USGS catalogs: (i) the Building Seismic Safety Council 2014 Event Set, BSSC14 (USGS, n.d.-a); (ii) the ShakeMap Scenario Catalog for Selected Quaternary Active Faults in Washington State, WA22 (USGS, n.d.-c); and (iii) the M9 Cascadia Earthquake Scenarios, CSZM9 (USGS, n.d.-b).

The BSSC14 is a large, national scenario catalog of deterministic ruptures used to generate the 2014 USGS National Seismic Hazard Map (Petersen et al., 2014). Here, we simulate 54 scenarios from BSSC14 which affect OR. The WA22 catalog includes known active or potentially active Quaternary faults in WA which may produce an earthquake large enough to damage structures, as identified by scientists at the USGS and Washington State Geological Survey. We simulate all 26 scenarios from WA22. We also simulate the 2$^{nd}$, 16$^{th}$, 50$^{th}$ (median), 84$^{th}$, and 98$^{th}$ percentile ground motions (i.e., 5 "scenarios") from the CSZM9 catalog. This catalog contains the ensemble ShakeMaps of Wirth et al. (2021) depicting thirty M9.0 Cascadia Subduction Zone (CSZ) ruptures as simulated by Frankel et al. (2018). While it is unlikely that an M9.0 CSZ rupture will produce ±1 or 2σ ground motions across the entire region in any one event, these ensemble ShakeMaps are intended for use in impact forecasting from extreme shaking on a localized scale. Additional details regarding scenario earthquake development for the BSSC14, WA22, and CSZM9 catalogs are provided on their respective webpages (USGS, n.d.-a; USGS, n.d.-c; USGS, n.d.-b).

To simulate the 85 scenarios, we employ the Sanger et al. (2024c) scripts in DesignSafe. These scripts interact with USGS ShakeMap ground motions (i.e., *PGA*) in .xml format, which are called via a user-input web address. The code converts the ShakeMap *PGA* to a raster and scales it to $PGA_M$. Given these inputs for each scenario, we surrogate the three geotechnical models (i.e., *LPI*, $LPI_{ISH}$, and *LSN*) and compute the corresponding probability of liquefaction-induced ground failure ("PGF") using the fragility functions of Geyin and Maurer (2020). PGF represents the median probability of observing liquefaction manifestation at the surface for any location within a map pixel. As such, PGF is a proxy of damage potential for near-surface infrastructure like roadways and is used as the damage metric in the hazard and impact forecasting examples presented in this manuscript. Finally, because the surrogated geotechnical models have different efficacies and it is unclear which model is best (e.g., Geyin et al., 2020; Rasanen et al., 2023), we average the three PGFs in each pixel for an "ensemble" model. The products and results of the liquefaction hazard forecasts for the 85 scenario simulations are discussed in *Results and Discussion*.

**Table 2.** Summary of fault rupture scenarios in the Pacific Northwest.

| Rupture Scenario | M | Catalog | Rupture Scenario | M | Catalog |
|---|---|---|---|---|---|
| Wallula fault zone (1936) | 6 | WA22 | McGee Mountain fault zone | 7 | BSSC14 |
| Yaquina faults | 6.1 | BSSC14 | Mount St. Helens Seismic zone | 7 | WA22 |
| Bolton fault | 6.2 | BSSC14 | Saddle Mountains fault | 7 | WA22 |
| Grant Butte fault | 6.2 | BSSC14 | South Klamath Lake West | 7 | BSSC14 |
| Eastern Pine Forest Range fault zone | 6.3 | BSSC14 | Warner Valley faults (west) | 7 | BSSC14 |
| Mount Hood fault | 6.3 | BSSC14 | Whaleshead fault zone | 7 | BSSC14 |
| Paulina Marsh fault | 6.3 | BSSC14 | Darrington Devils fault | 7.1 | WA22 |
| South Slough thrust and reverse faults | 6.3 | BSSC14 | Entiat Fault | 7.1 | WA22 |
| Helvetia fault | 6.4 | BSSC14 | Klamath graben fault system (west) | 7.1 | BSSC14 |
| Waldport fault | 6.4 | BSSC14 | Manastash fault | 7.1 | WA22 |
| Beaver Creek fault zone | 6.5 | BSSC14 | Portland Hills fault | 7.1 | BSSC14 |
| Juniper Mountain fault | 6.5 | BSSC14 | Sky Lakes fault zone | 7.1 | BSSC14 |
| Sandy River fault zone | 6.5 | BSSC14 | Tacoma fault | 7.1 | WA22 |
| Boulder Creek fault | 6.6 | WA22 | Wallowa fault | 7.1 | BSSC14 |
| Happy Camp fault | 6.6 | BSSC14 | Wallula fault zone | 7.1 | WA22 |
| Pine Valley graben fault system, Brownlee sec | 6.6 | BSSC14 | Alvin Canyon fault | 7.2 | BSSC14 |
| Turner and Mill Creek faults | 6.6 | BSSC14 | Chemult graben fault system (east) | 7.2 | BSSC14 |
| Davis Creek | 6.7 | BSSC14 | Chemult graben fault system (west) | 7.2 | BSSC14 |
| Lacamas Lake fault | 6.7 | BSSC14 | Gales Creek fault | 7.2 | WA22 |
| Lake Creek - Boundary Creek - Sadie Creek fault | 6.7 | WA22 | Horse Heaven Hills fault | 7.2 | WA22 |
| Utsalady Point fault (east) | 6.7 | WA22 | South Whidby Island Fault | 7.2 | WA22 |
| Utsalady Point fault (west) | 6.7 | WA22 | Southeast Newberry fault zone | 7.2 | BSSC14 |
| Coquille anticline | 6.8 | BSSC14 | Spencer Canyon fault | 7.2 | WA22 |
| Eastern Bilk Creek Mountains fault zone | 6.8 | BSSC14 | Winter Rim fault system | 7.2 | BSSC14 |
| Gales Creek fault zone | 6.8 | BSSC14 | Daisy Bank fault | 7.3 | BSSC14 |
| Gillem-Big Crack | 6.8 | BSSC14 | Hite fault system | 7.3 | BSSC14 |
| Latah fault | 6.8 | WA22 | Wecoma fault | 7.3 | BSSC14 |
| Mount Angel fault | 6.8 | BSSC14 | Hoppin Peaks fault zone | 7.4 | BSSC14 |
| Rush Peak fault | 6.8 | BSSC14 | Klamath graben fault system (east) | 7.4 | BSSC14 |
| Stonewall anticline | 6.8 | BSSC14 | Metolius fault zone | 7.4 | BSSC14 |
| Toe Jam fault | 6.8 | WA22 | Santa Rosa fault system | 7.5 | BSSC14 |
| Birch Bay fault | 6.9 | WA22 | Seattle fault zone | 7.5 | WA22 |
| Newberg fault | 6.9 | BSSC14 | Steens fault zone | 7.5 | BSSC14 |
| Pine Valley graben fault system, Halfway-Pose | 6.9 | BSSC14 | Topponish Ridge - Mill Creek fault | 7.5 | WA22 |
| Rattlesnake Mountain fault | 6.9 | WA22 | Warner Valley faults (east) | 7.5 | BSSC14 |
| Tule Springs Rim Fault | 6.9 | BSSC14 | Seattle fault zone/Saddle Mountain (east) | 7.8 | WA22 |
| Abert Rim fault | 7 | BSSC14 | Big Lagoon-Bald Mtn | 7.9 | BSSC14 |
| Battle Rock fault zone | 7 | BSSC14 | Cascadia, 2$^{nd}$ percentile | 9 | CSZM9 |
| Canyon River - Saddle Mountain (east) | 7 | WA22 | Cascadia, 16$^{th}$ percentile | 9 | CSZM9 |
| Canyon River - Saddle Mountain (middle) | 7 | WA22 | Cascadia, 50$^{th}$ percentile | 9 | CSZM9 |
| Canyon River - Saddle Mountain (west) | 7 | WA22 | Cascadia, 84$^{th}$ percentile | 9 | CSZM9 |
| Cape Blanco anticline | 7 | BSSC14 | Cascadia, 98$^{th}$ percentile | 9 | CSZM9 |
| Cottonwood Mountain fault | 7 | BSSC14 | | | |

## Results and Discussion

### *Model Performance in the Pacific Northwest*

To understand model performance in the region of study, we benchmark the GLMs against the state-of-practice geospatial liquefaction model of Rashidian and Baise (2020), henceforth referred to as "RB20." There are several variants of this model in use (e.g., Zhu et al. 2017; Rashidian and Baise 2020; Allstadt et al., 2022), but the results reported here reflect predictions made by the RB20 variant implemented in the USGS ground failure product (Allstadt et al., 2022), as this model performs best in the presented tests. The RB20 model coefficients and implementation details employed in this comparison are summarized in the *Supplemental Information*. We are not aware of any other geospatial liquefaction models that can be deployed and compared in this region.

For the purposes of this study, it would be most interesting to compare the model performances in the Pacific Northwest. However, due to the previously described seismic setting of the region, with long return periods for moderate-to-large magnitude earthquakes, there is only one significant recorded event on which we can test: the 28 February 2001 M6.8 Nisqually earthquake. We test the GLMs and RB20 using the case histories compiled from this event by Rasanen et al. (2022, 2023), of which there are nine (9) positive observations (i.e., liquefaction observed) and 15 negative observations. This, of course, is an exceedingly small case history set, and it is not an unbiased test. Liquefaction manifestation observations made during the Nisqually earthquake by Bray et al. (2001) are included in the RB20 training data and the Sanger et al. (2025) GLMs trained on CPTs from the affected region. Therefore, in addition to this test, we present summary statistics from Sanger et al. (2025), who tested the GLMs and RB20 in regions without explicit data in either model's training set. Here, three liquefaction inventories were compiled for a global "unseen" test that (a) postdate RB20's training and (b) occurred in regions where no CPTs were compiled in the GLM training data: the 2019 Puerto Rico (Allstadt and Thompson, 2021), 2019 Ridgecrest (Zimmaro et al., 2020), and 2023 Turkey earthquakes (Cetin et al., 2023; Taftsoglou et al., 2023). In the global "unseen" tests, negative observations were randomly sampled from map cells without positive observations to balance the class set. The results of the Nisqually and global "unseen" test are summarized in Table 3.

To quantify model performance, we employ the Brier Score (*BS*):

$$Brier\ Score\ (BS) = \frac{1}{N}\sum_{i=1}^{N}(P_i - O_i)^2 \qquad \text{(Eq. 6)}$$

where *N* is the number of observations, *P* is the predicted probability, *O* is the observed probability (0 or 1), and *i* is the observation index. For performance testing on liquefaction case history observations, we deploy the GLM with the fragility functions of Geyin and Maurer (2020), such that the resultant predictions are probability of liquefaction-induced ground failure (i.e., ejecta or surface deformation), previously defined as "PGF." *BS* ranges from 0 (perfect model) to 0.5 (random guess) to 1 (perfectly incorrect model), with 0.25 representing uniform 50% predictions. A *BS* below 0.25 reflects an increasingly "good" model as *BS* approaches 0. The *BS* captures both class separation and calibration near 50% probability, offering advantages over metrics like area under a receiver-operating-characteristic curve (*AUC*) (e.g., Fawcett, 2006), which reflects class separation but not calibration. To assess whether differences in model performance are statistically significant, the *BS* test results are bootstrap sampled 10,000 times to compute *BS* confidence intervals (CIs), and spatial autocorrelation is accounted for using the Moran's *I* statistic (Anselin, 1996). For tests with $I > 0.3$, the *BS* distribution is resampled via agglomerative clustering (e.g., Steinbach, 2000), which intends to remove spatial correlation. The bootstrapped and spatially stratified performance arrays are further evaluated for statistically significant differences using the Kolmogorov-Smirnov (KS) two-sample test (Smirnov, 1939), and Cohen's *d* effect (Cohen, 1988). The two-sample KS test quantifies the maximum absolute difference between two cumulative distribution functions, where KS test statistics are interpreted as: 0-0.2, little to no difference; 0.2-0.5, moderate differences; and greater than 0.5, significantly different distributions. Furthermore, Cohen's *d* quantifies the magnitude and direction of the difference between two distributions, where *d* indicates the scale of difference in terms of standard deviations (i.e., *d* of 1 denotes distributions differ by one standard deviation), and the sign indicates the direction of the difference (i.e., a negative value denotes an improvement of the GLMs over RB20).

In the global "unseen" case histories, Sanger et al. (2025) observed all GLM models (*LPI*, $LPI_{ISH}$, *LSN*, and the averaged "ensemble" model) outperformed RB20 to an apparently significant degree, as indicated by 99% CIs and KS test statistics on the *BS* distributions, wherein Cohen's *d* suggests that the GLM improved upon the *BS* of RB20 by about 1.5 standard deviations. The limited testing in the Pacific Northwest using the Nisqually case history observations suggests similar trends, where the mean *BS* of all GLM models is less than that of RB20, and Cohen's *d* suggests that the GLMs improve upon the *BS* of RB20 by about one standard deviation. However, given the small, imbalanced test set, these results are interpreted as generally informative but not conclusive with respect to statistical significance. Interested readers are referred to Sanger et al. (2025) for other testing scenarios and otherwise invited to test further.

**Table 3.** Summary of GLM testing performance compared to RB20.

| Case Histories | Model | Mean *BS* | 99% Confidence Interval of Mean *BS* | Comparison Against RB20 | |
|---|---|---|---|---|---|
| | | | | KS Test Statistic | Cohen's *d* Effect |
| Sanger et al. (2025) Global "unseen" test | RB20 | 0.393 | 0.380 - 0.407 | - | - |
| | *LPI*-ML | 0.153 | 0.143 - 0.162 | 0.56 | -1.51 |
| | $LPI_{ISH}$-ML | 0.128 | 0.120 - 0.137 | 0.62 | -1.62 |
| | *LSN*-ML | 0.180 | 0.170 - 0.191 | 0.49 | -1.46 |
| | Ensemble | 0.146 | 0.138 - 0.155 | 0.57 | -1.59 |
| Nisqually | RB20 | 0.168 | 0.158 - 0.181 | - | - |
| | *LPI*-ML | 0.139 | 0.069 - 0.234 | 0.30 | -0.962 |
| | $LPI_{ISH}$-ML | 0.160 | 0.151 - 0.170 | 0.31 | -1.093 |
| | *LSN*-ML | 0.122 | 0.112 - 0.133 | 0.26 | -1.122 |
| | Ensemble | 0.159 | 0.085 - 0.256 | 0.19 | -0.990 |

***Liquefaction Hazard and Impact Forecasting for Scenario Earthquakes***

The resulting liquefaction hazard forecasts for all 85 scenario earthquakes are provided in digital format as a DesignSafe data repository (Sanger and Maurer, 2025). Example products are shown in Fig. 3 for a M7.1 rupture of the Portland Hills Fault, which is a major crustal-fault hazard for the city of Portland, OR. The results include (i) a map and raster of *PGA* from the USGS ShakeMap, reinterpreted to the desired resolution (Fig. 3a); (ii) a $PGA_M$ raster; (iii) maps and rasters of the *LPI*, $LPI_{ISH}$, and *LSN* predicted for the event (Fig. 3d-f) with (iv) maps and rasters of the corresponding PGFs according to the Geyin and Maurer (2020) fragility functions (Fig. 3g-i); (v) an ensembled PGF map and raster, computed as the spatial average PGF from the three surrogated geotechnical models (Fig. 3b); and (vi) a simplified classification measure of variance in ML model predictions, which illustrates the extent to which the map is updated by in-situ geotechnical data (i.e., locally updated) (Fig. 3c). As described in Sanger et al. (2025), the geotechnical influence maps are intended to communicate where, and to what degree, the predicted response is driven by measured geotechnical data and models. In these maps, the variance of the kriged residuals approaches zero at CPT sites such that predictions are based mostly on known subsurface conditions and have lower model uncertainty (i.e., the geotechnical influence is major), and the variance increases toward the overall model uncertainty at locations distant from CPTs (i.e., there is no geotechnical influence). The results shown in Fig. 3 are provided for each scenario and are organized whereby each scenario has a subdirectory containing maps of the results (PDFs), reference raster data (GeoTIFFs), and a corresponding ArcGIS Map package.

The results mapped in Fig. 3 and available for 85 scenario events have broad functionality for disaster preparedness, land-use planning, infrastructure impact analysis, and other regional-sale uses. For example, planners and emergency managers in Portland can use these maps to identify neighborhoods and corridors where the scenario M7.1 rupture of the Portland Hills Fault produces high predicted liquefaction hazard (Fig. 4). As could be expected, the hazard pattern shows a strong spatial correlation with the confluence of the Columbia and Willamette Rivers, which produce thick deposits of saturated alluvial deposits. In this case, high PGF values are concentrated along the waterfront districts of Portland (e.g., the Pearl District). These areas host dense urban development and critical port, transportation, medical, and lifeline infrastructure, underscoring the importance of emergency planning. The elevated liquefaction hazard in this corridor is driven by both intense ground motions near the fault rupture and geologic conditions that promote high susceptibility, as captured by the GLMs. Beyond individual scenarios, the ensemble of hazard forecasts offers a portfolio of plausible events for probabilistic evaluations of urban resilience. In this way, the GLMs serve not only as scientific outputs but also as actionable tools for decision-making. In the following, several other scenario earthquakes of particular regional interest are discussed, after which some examples of downstream impact analysis are given.

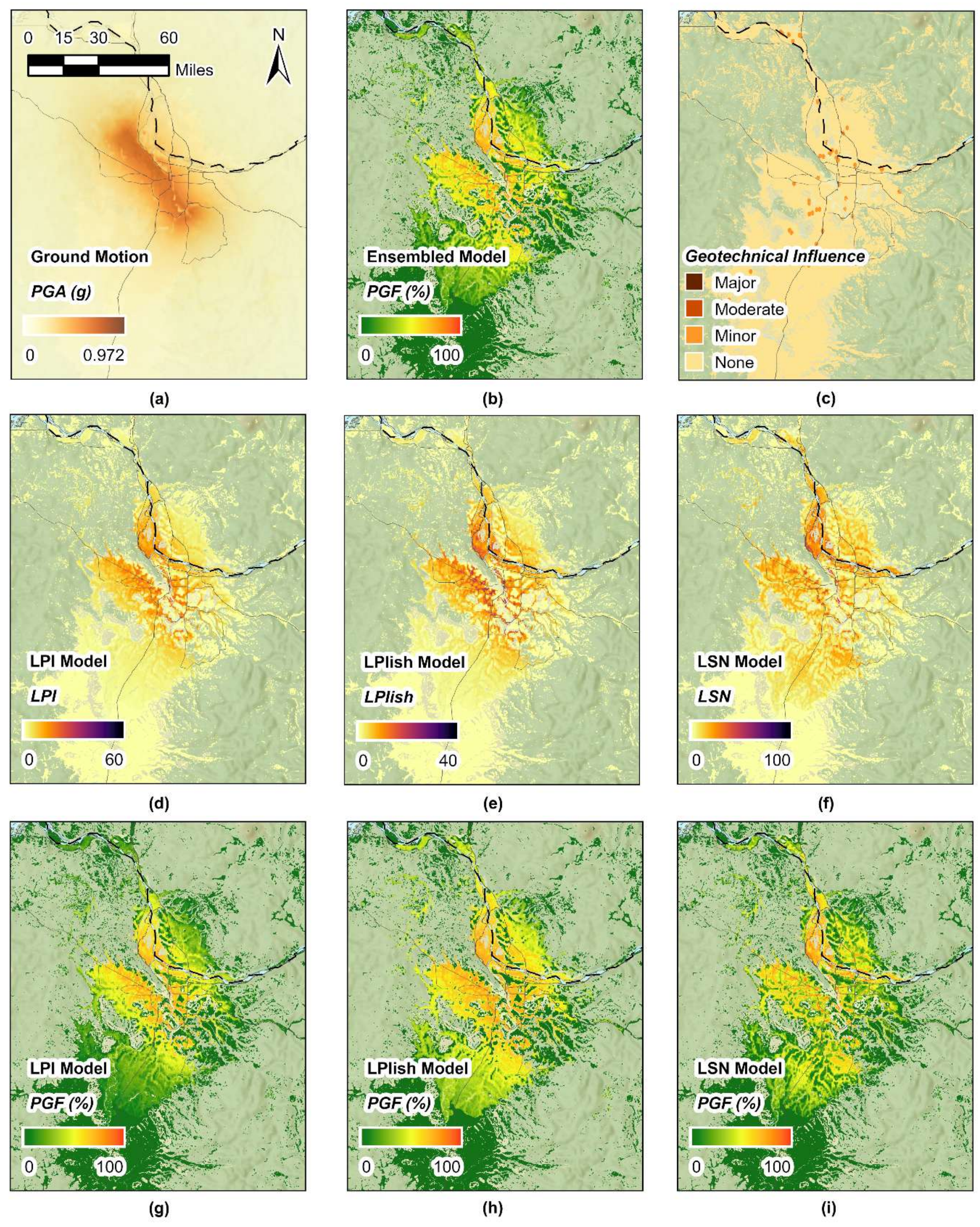

**Fig. 3.** Example products for a scenario M7.1 rupture of the Portland Hills Fault, including (a) *PGA*; (b) ensembled PGF; (c) geotechnical influence on predictions; (d) *LPI*; (e) $LPI_{ISH}$; (f) *LSN*; (g) PGF based on *LPI*; (h) PGF based on $LPI_{ISH}$; and (i) PGF based on *LSN*. (Sources: Esri and USGS.)

Other municipalities in the region would benefit similarly from interrogating the predicted liquefaction hazard resulting from ruptures of local crustal faults. For example, the scenario M7.2 rupture of the South Whidbey Island Fault in WA, which is a major crustal fault hazard for the Puget Sound region, predicts high PGFs concentrated along the Snohomish River corridor between Everett and Snohomish (Fig. 5). This pattern reflects the presence of thick, Holocene-age alluvium and coastal estuarine deposits combined with shallow groundwater. Notably, Interstate-5, the only north-south interstate on the U.S. west coast, runs through this high-hazard area; moreover, numerous large paleoliquefaction features from prehistoric earthquakes have been documented near Interstate-5 bridges and roadway in the Snohomish River delta (e.g., Rasanen et al., 2021), lending credence to the modeling results. Elevated PGF values are also observed on the southern ends of Whidbey and Camano Islands, where similar sediment deposits are common, albeit on a smaller scale, and could potentially impact ports and harbors, including passenger ferry terminals.

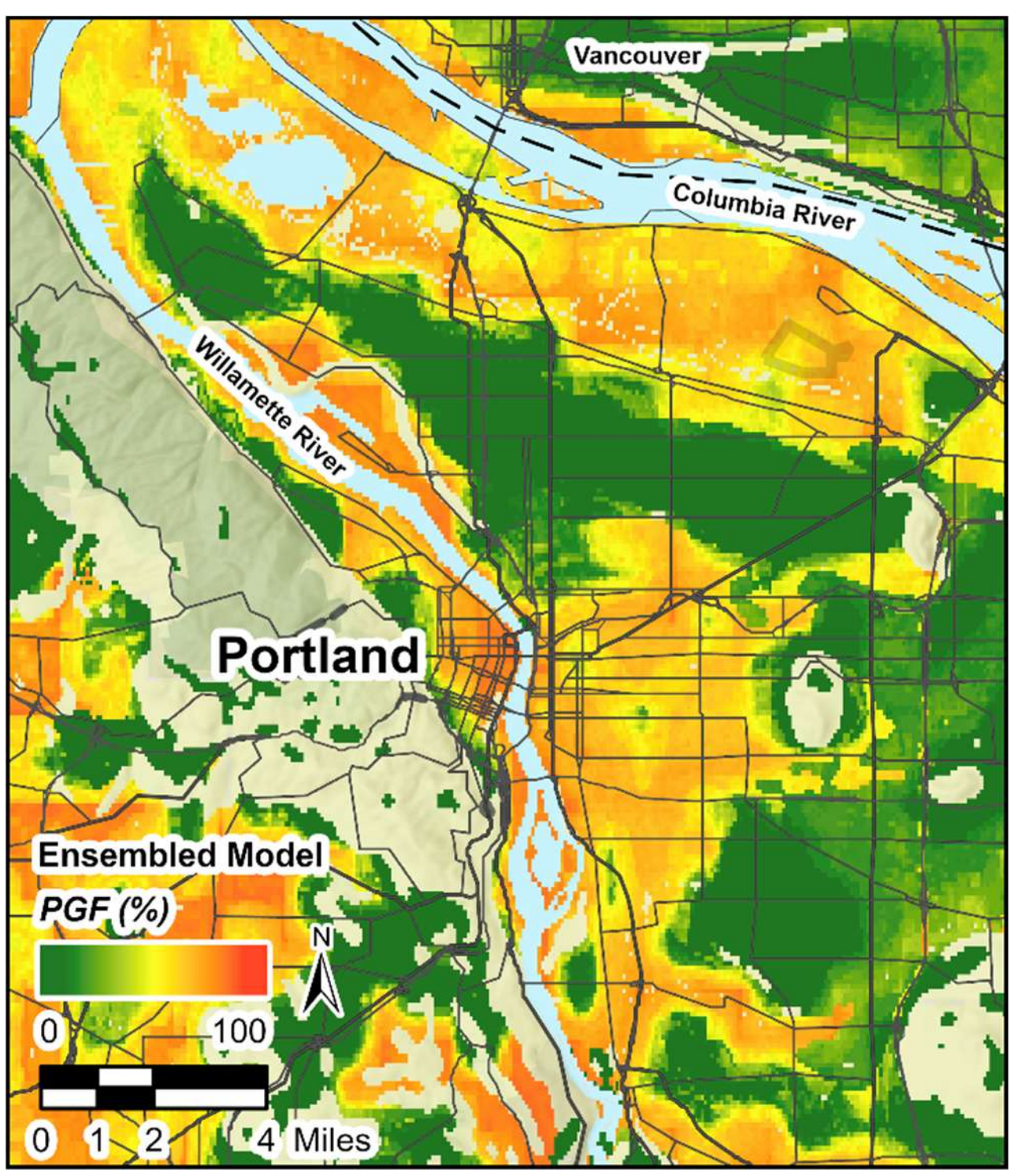

**Fig. 4.** Predicted PGF for a scenario M7.1 rupture of the Portland Hills Fault in Portland, OR. (Sources: Esri and USGS.)

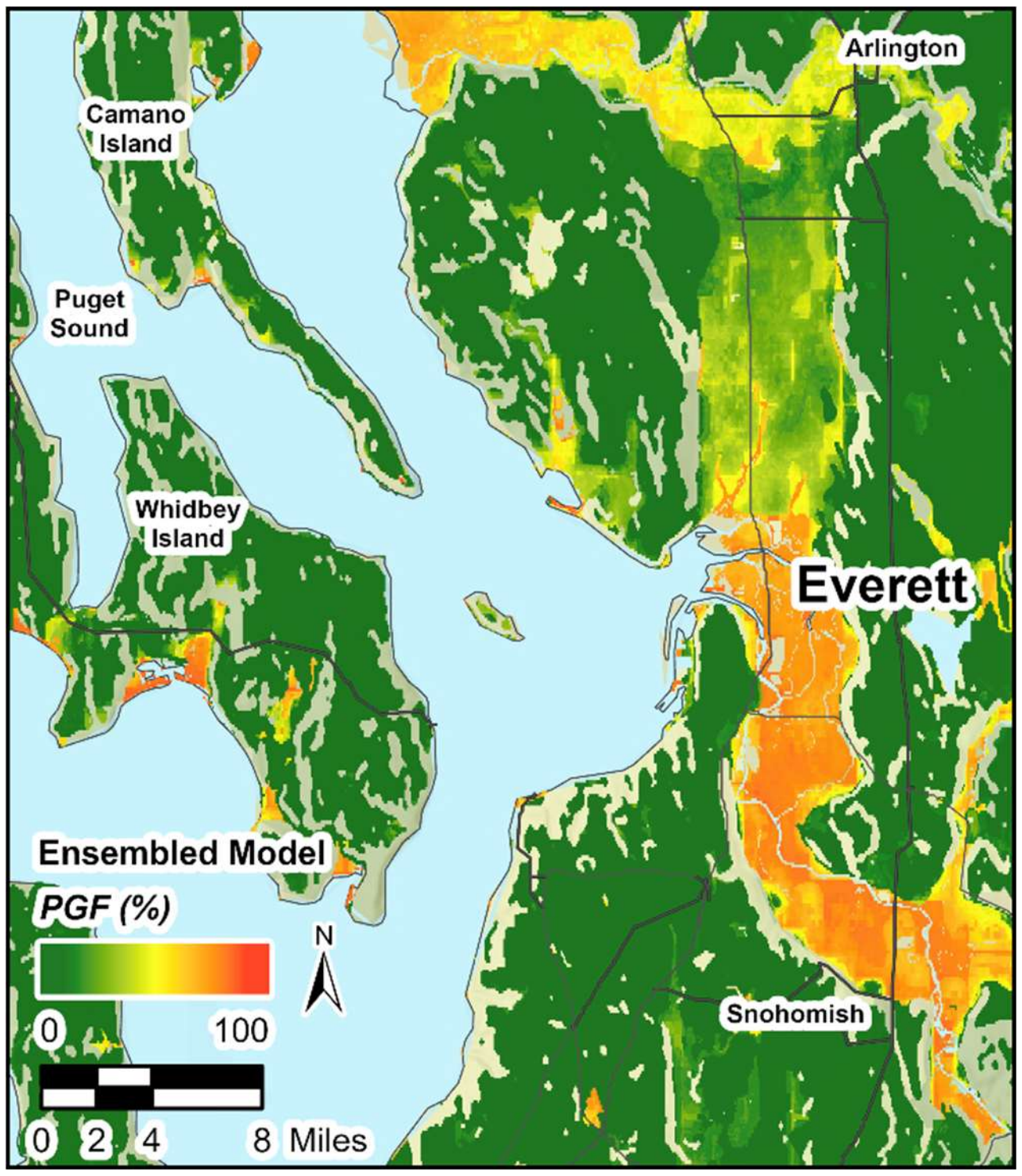

**Fig. 5.** Predicted PGF for a scenario M7.2 rupture of the South Whidbey Island Fault near Everett, WA. (Sources: Esri and USGS.)

Another major crustal-fault hazard – the Seattle Fault Zone – lays 30 km to the south and is believed to be capable of producing an M7.5 rupture. In Seattle, predicted PGFs for this event are notably high in areas adjacent to Elliott Bay and along the Duwamish River corridor, extending from the International District southward through the industrial zones of Georgetown and Tukwila (Fig. 6). Additional zones of heightened hazard appear along Lake Washington's western shoreline, particularly near Rainier Beach, where saturated fill and deltaic deposits are common. As with the prior scenarios, the high PGFs in Seattle coincide with dense port, utility, and transportation infrastructure, and thus have heightened potential to disrupt the largest seaport in the U.S. Pacific Northwest, through either direct damage to port infrastructure, or to the road, rail, or utility networks that serve it. In contrast, areas of higher elevation throughout Seattle (e.g., Queen Anne, Capitol Hill) are typically comprised of glacial tills and exhibit negligible hazard.

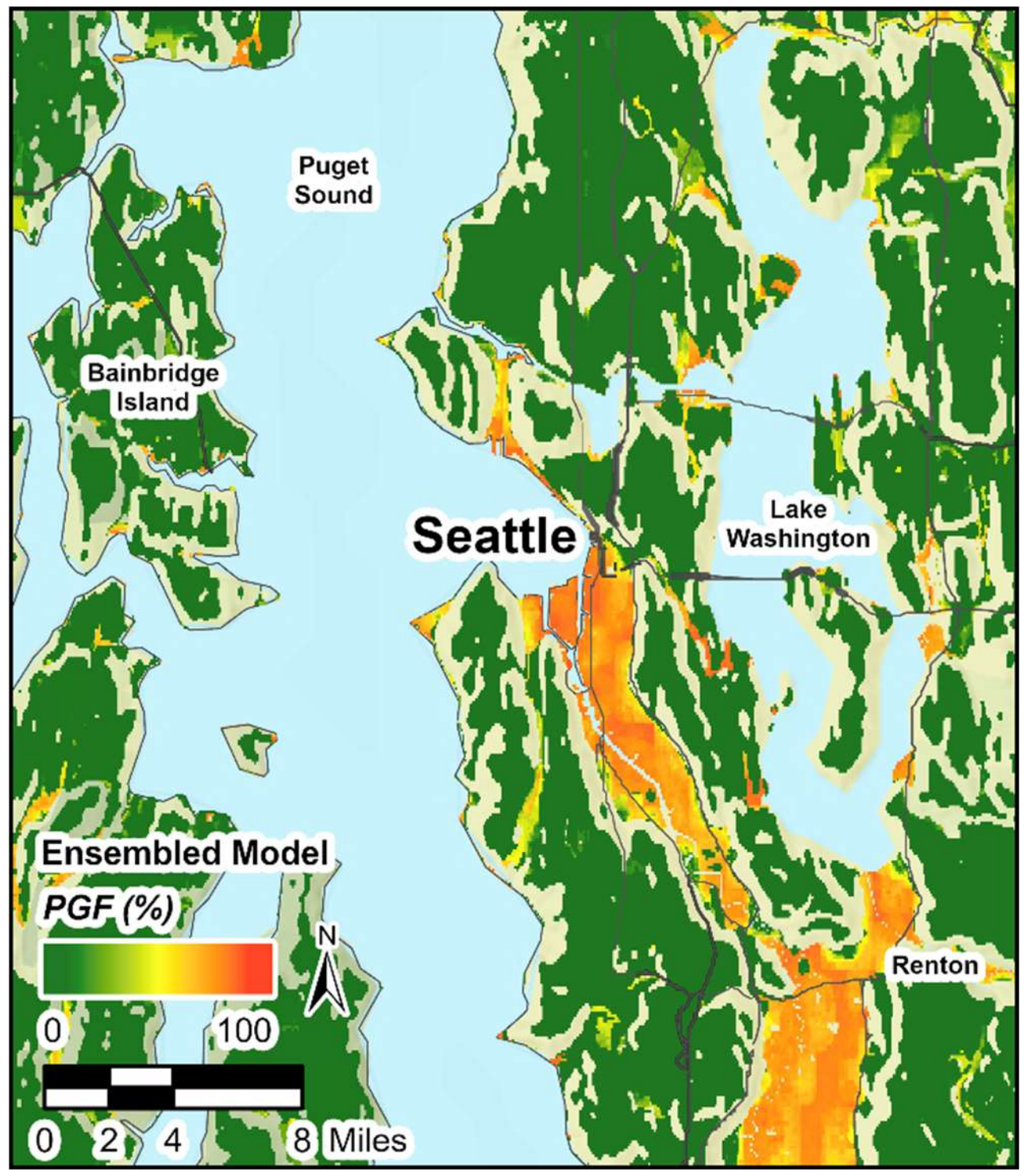

**Fig. 6.** Predicted PGF for a scenario M7.5 rupture of the Seattle Fault Zone in Seattle, WA. (Sources: Esri and USGS.)

Beyond crustal faults, the CSZ poses a major hazard to the entire region, with high potential for liquefaction across hundreds of square kilometers stretching from Northern California to Southern British Columbia, and as far east as the Cascade Range. Predicted PGFs considering an M9.0 CSZ rupture (median scenario), are shown in Fig. 7 for the Portland, OR and Seattle, WA metropolitan areas. The expected PGFs follow very similar patterns as compared to the respective crustal-fault scenarios in these cities, although the PGFs tend to be lower in the CSZ scenario, which is attributable to the large site-to-source distance between CSZ ruptures and the Puget and Willamette lowlands. Fig. 8 quantifies the differences in predicted PGFs between the crustal and subduction-zone events for each of these cities. In the urban cores of Portland and Seattle, PGFs are up to 25% lower in CSZ events. Thus, although CSZ hazards rightfully receive considerable media attention and greater public awareness, ground failure is more likely in the crustal events, at least on a local scale. Of course, what may be missed looking only at Portland and Seattle in isolation is the immensity of the area affected by CSZ ruptures and the potential for ground-failure simultaneously in numerous regional hubs of population and commerce.

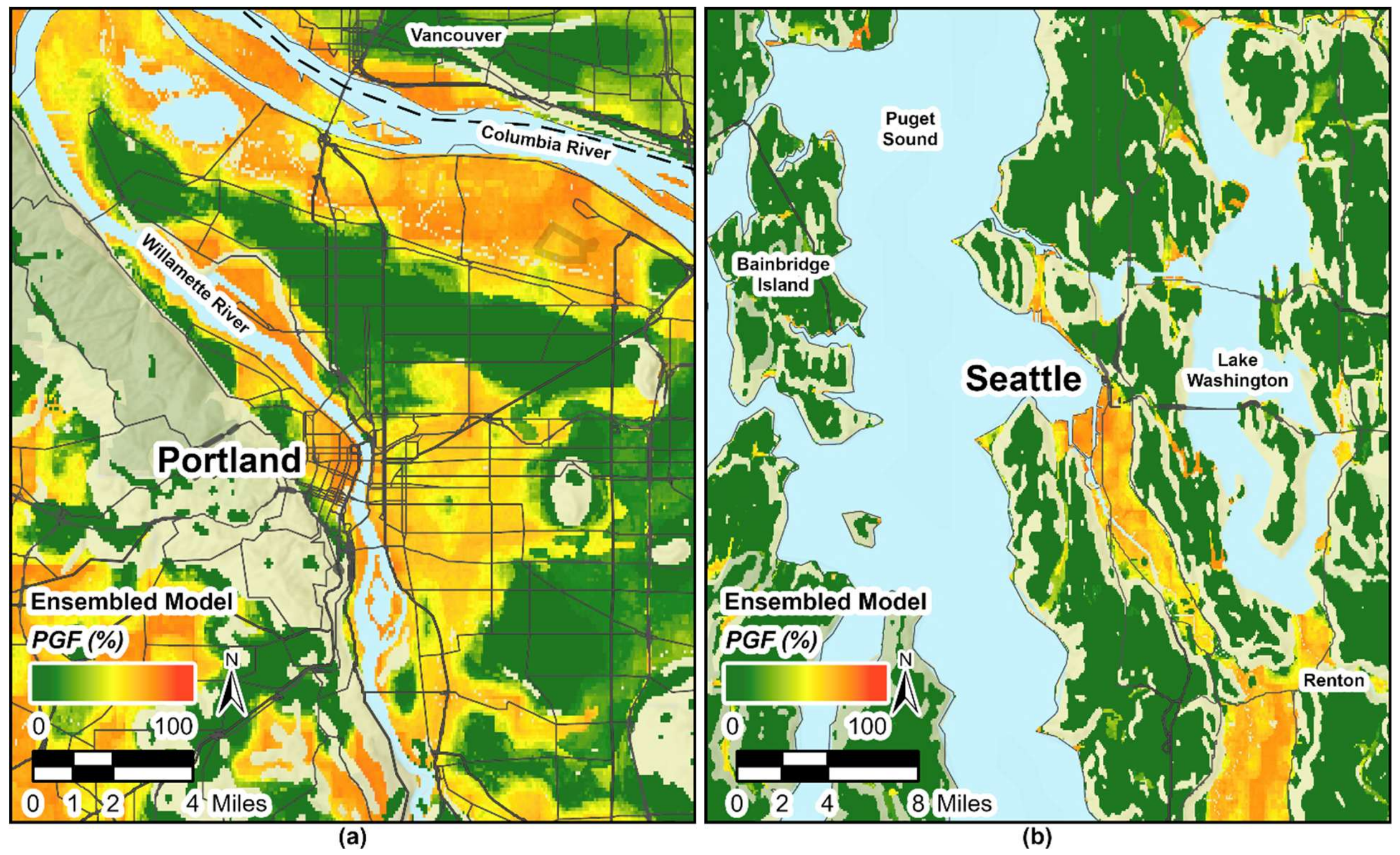

**Fig. 7.** Predicted PGF for a scenario M9.0 rupture of the CSZ (median scenario) near (a) Portland, OR and (b) Seattle, WA. (Sources: Esri and USGS.)

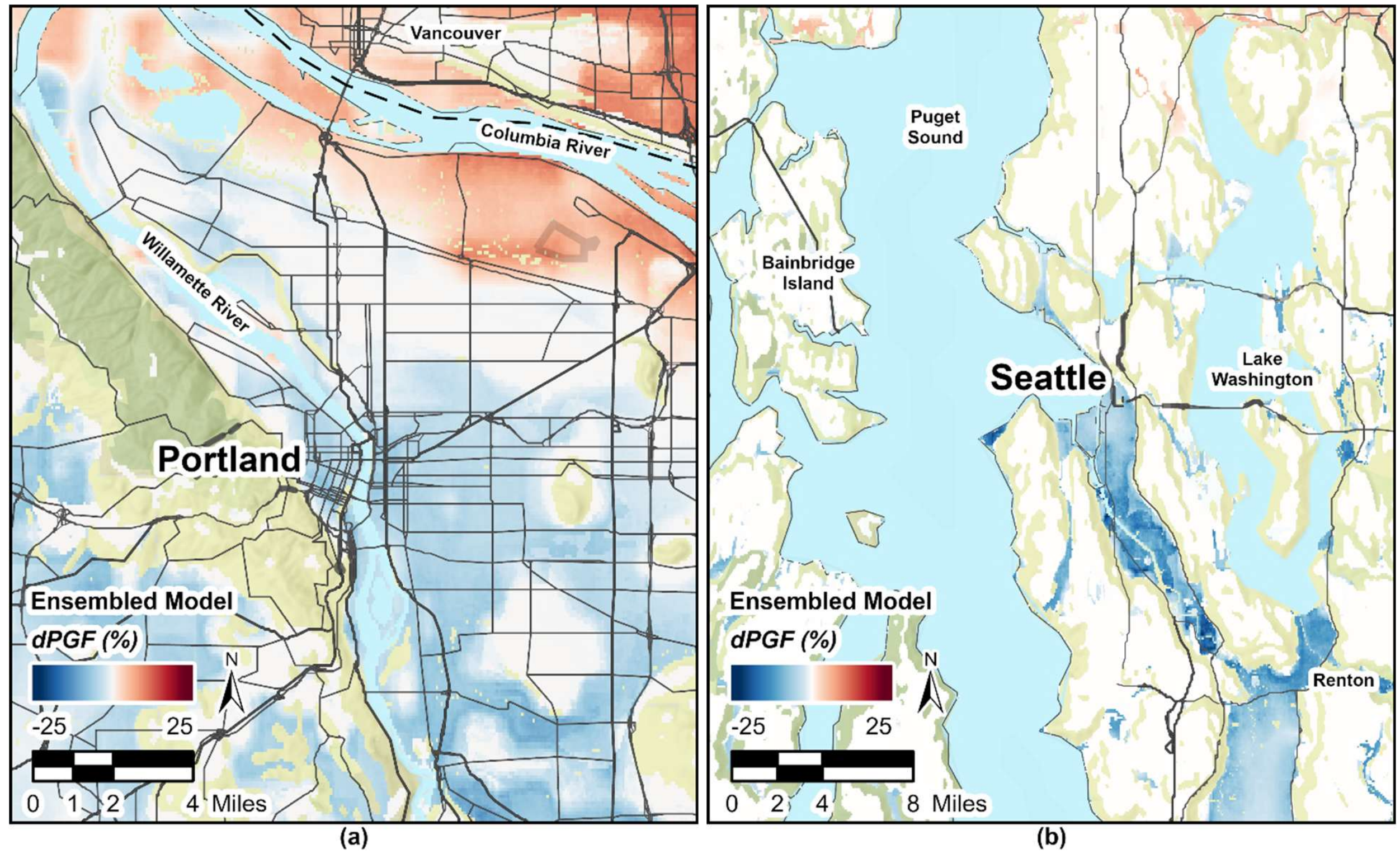

**Fig. 8.** Difference in predicted PGFs between a M9.0 rupture of the CSZ (median scenario) and (a) a M7.1 rupture of the Portland Hills Fault in Portland, OR and (b) a M7.5 rupture of the Seattle Fault Zone in Seattle, WA. Blue indicates higher hazard predicted for the local, crustal scenario, and red indicates higher hazard for the M9.0 scenario. (Sources: Esri and USGS.)

To explore how GLM outputs could be extended to downstream consequences, we first demonstrate impacts to distributed infrastructure, such as pipelines or road networks. A municipality such as Bandon, OR, for example, may consider how liquefaction damage to roads and bridges could impact tsunami evacuation planning. Fig. 9 forecasts PGFs at the municipality scale for a scenario M9.0 CSZ rupture affecting Bandon. Panel (a) shows the predicted PGF, with spatial variation reflecting both the ground motion distribution and site-specific liquefaction susceptibility. Notably, several areas near the mouth of the Coquille River and along low-lying coastal zones exhibit elevated PGFs, suggesting a heightened risk of ground deformation. Panel (b) overlays the PGF predictions onto the public road network (ODOT, 2025), enabling direct assessment of potential disruptions during tsunami evacuations. This type of analysis supports actionable, scenario-based vulnerability assessments. For example, Bandon officials can evaluate whether designated evacuation routes intersect with high-PGF zones and prioritize those for mitigation or alternate routing. The ability to map and quantify such impacts before an event occurs underscores the practical utility of the GLM in integrated emergency management and resilient infrastructure planning.

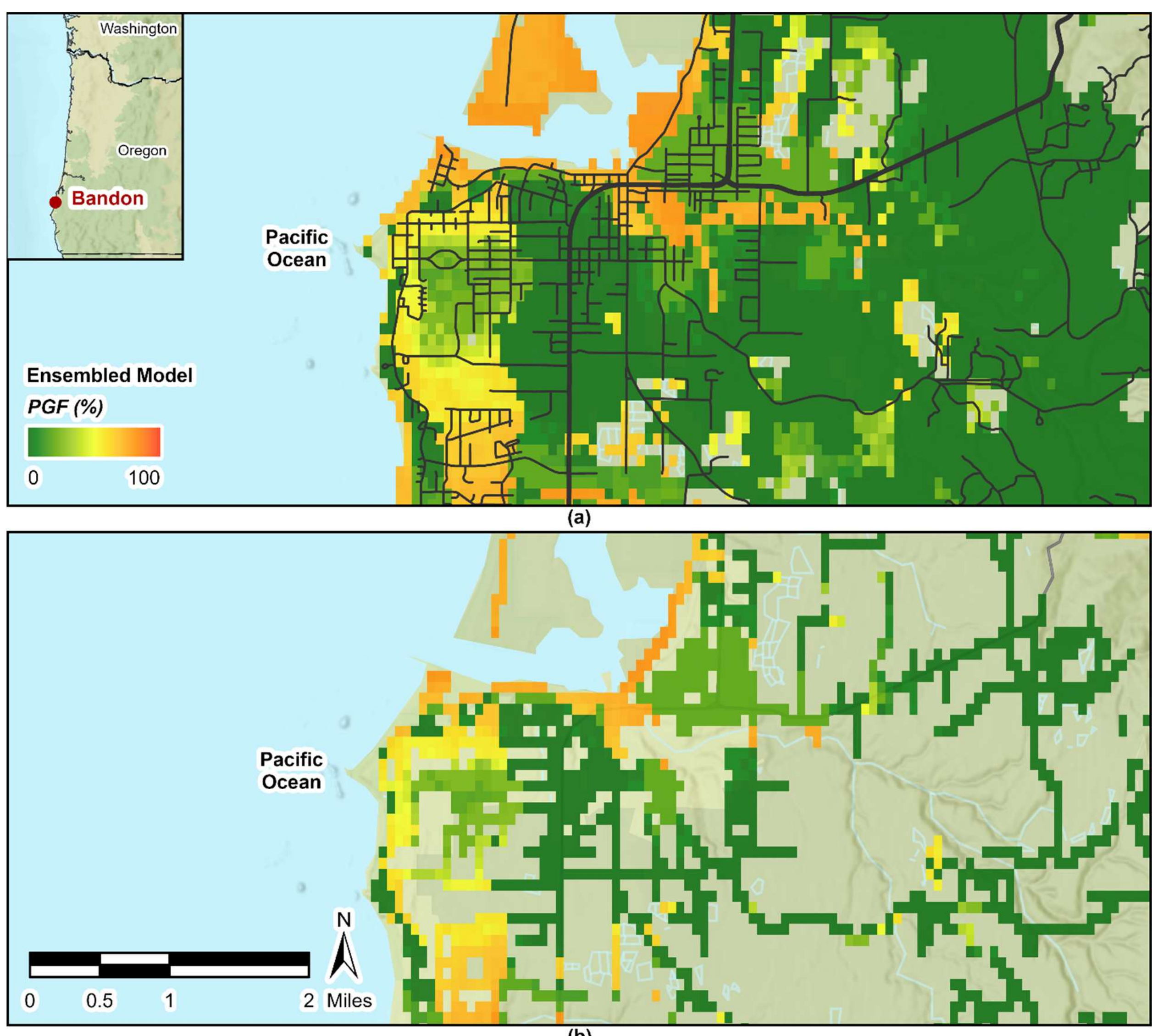

**Fig. 9.** Predicted PGF in Bandon, OR for an M9.0 CSZ rupture (median scenario), shown (a) at the municipality scale with (b) consideration of impacts on the public road network. (Sources: Esri and USGS.)

Additionally, managers and owners of distributed assets such as bridges or dams may consider the regional vulnerability of their asset portfolio to inform mitigation priorities or make other planning decisions. Figs. 10 and 11 respectively show predicted PGFs at earth dam sites (USACE, 2024) and arterial bridge sites (USDOT; FHWA; BTS, 2020) in western WA for an M9.0 CSZ rupture (median scenario). The spatial distribution of PGF highlights that many dam and most especially, bridge sites—particularly those in estuarine environments and low-lying alluvial plains—are susceptible to moderate to high liquefaction hazard during a large CSZ event. These predictions enable dam and bridge owners, regulators, and emergency managers to identify facilities that may be at risk of foundation instability, lateral spreading, deformation of dam embankments, or disruption to associated lifelines such as access roads and pipelines. Notably, these predictions do not account for dam or bridge type, geometry, engineering design, etc., or for site-specific liquefaction mitigation that may have been performed. Indeed, most large dams and bridges are carefully engineered and strictly regulated with geohazards in mind. Yet smaller dams and bridges may predate rigorous engineering design – particularly with respect to liquefaction hazards – and are often less regulated. Many states do not require routine inspections or do not apply stringent safety standards to dams below certain height or storage thresholds, for example. In this regard, the GLM outputs provide a means for assessing regional dam and bridge exposure and prioritizing site-specific studies based on first-order screening.

Fig. 12 illustrates population impacts in an M9.0 scenario and speaks to potential for multiple large cities to be impacted simultaneously. As population density increases – according to CIESIN (2017) – the median PGF increases, meaning that the most populated areas are also the areas most likely to be damaged by liquefaction. This is not necessarily surprising, as cities tend to be in low-laying coastal and alluvial settings. Nonetheless, it is striking to observe that across all map pixels with a population density exceeding 1000 persons/km$^2$, the median PGF is nearly 40%. In total, more than 4 million people across the region are exposed to predicted PGFs greater than 50% in a median M9.0 scenario.

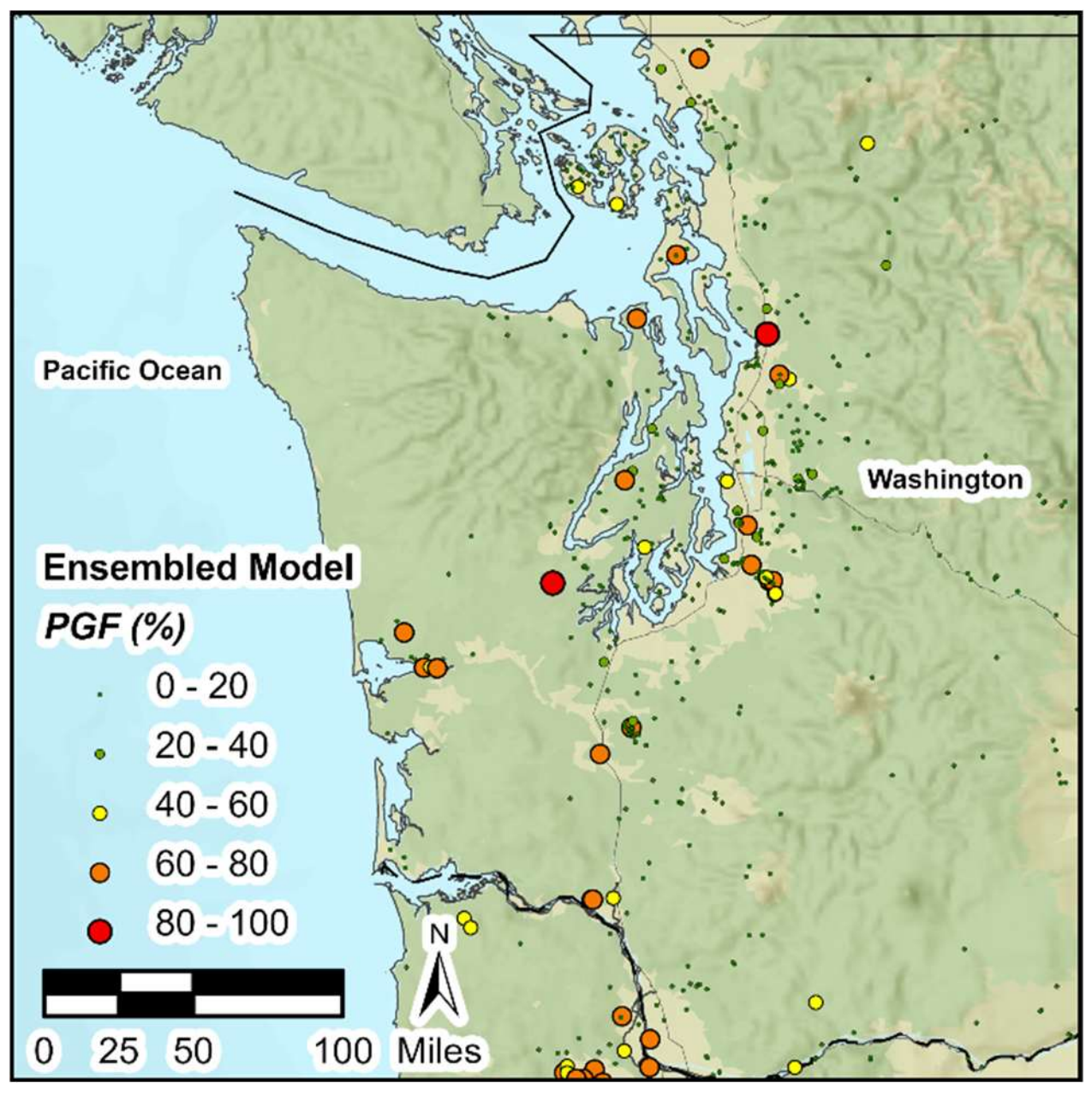

**Fig. 10.** Predicted PGF at earth dam sites in western WA given an M9.0 CSZ rupture (median scenario). (Sources: Esri and USGS.)

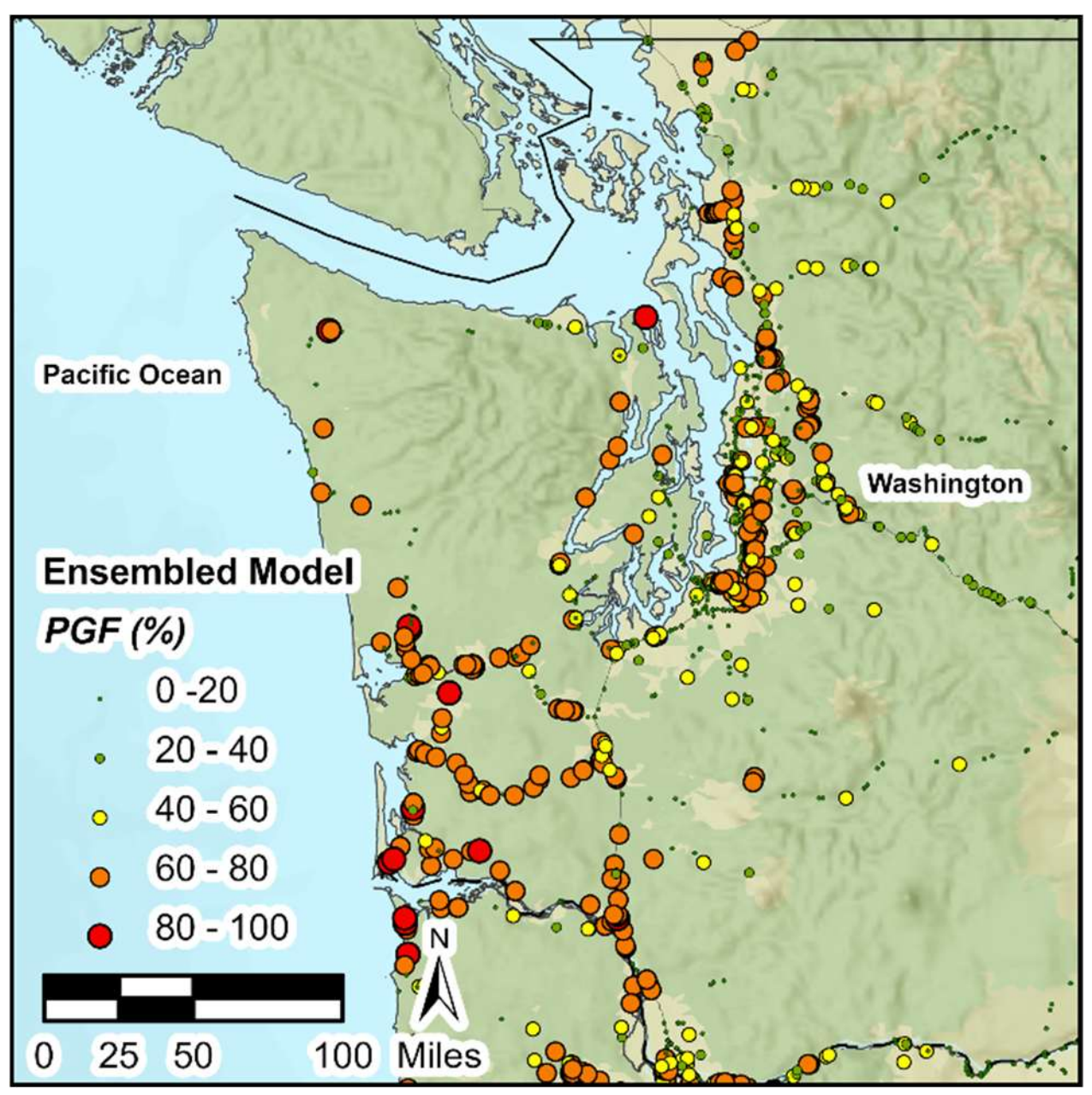

**Fig. 11.** Predicted PGF at arterial bridge locations in western WA given an M9.0 CSZ rupture (median scenario). (Sources: Esri and USGS.)

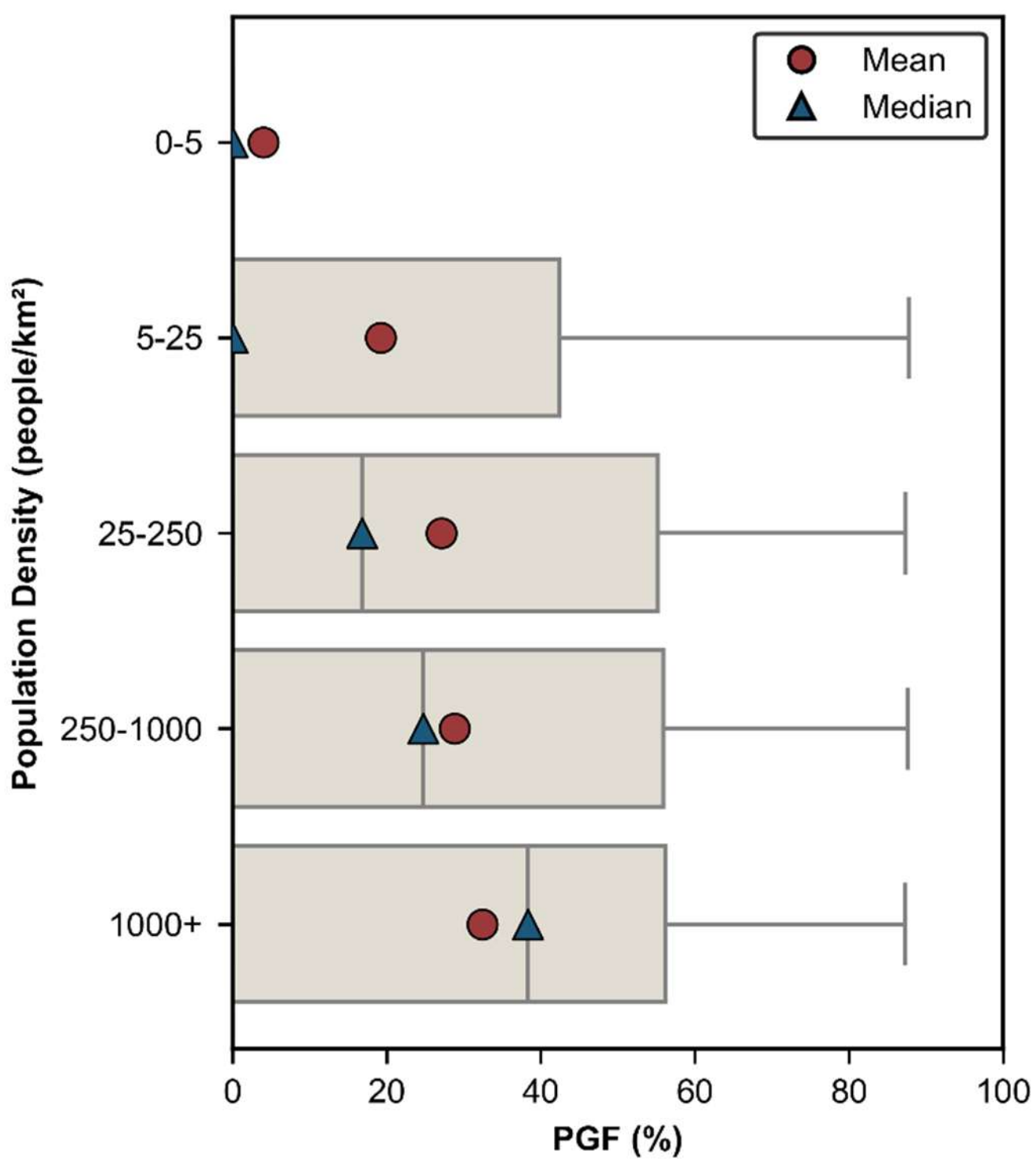

**Fig. 12.** Distribution of predicted PGF binned by population density in an M9.0 CSZ rupture. The box represents the interquartile range (IQR) from the first quartile, Q1 (i.e., 25th percentile) to the third quartile, Q3 (i.e., 75th percentile). The whisker represents Q3+1.5*IQR, given that Q1 is 0 for all bins.

***Limitations and Uncertainties***

The presented liquefaction hazard simulations and impact forecasts are subject to limitations and uncertainties not yet discussed. *First,* as surrogate models, these GLMs should be superseded at specific locations by site-specific, measured geotechnical measurements and analysis. This is, in effect, the aim of the geostatistical updating procedure, but site-specific data and analyses not included in the GLMs can be leveraged via further geostatistical updating. Further, there are several aspects and sources of uncertainty in the geotechnical models which propagate through to the geospatial surrogates of those models. Even ML models which precisely reproduce geotechnical models at all locations would not guarantee exact predictions of liquefaction phenomena. Recent tests of geotechnical triggering and manifestation models on global case histories demonstrate an approximate median *AUC* of 0.78 (Geyin et al., 2020). These models are known, for example, to be less constrained in regions of their parameter spaces with few training data, such as large-magnitude subduction zone earthquakes, and to potentially be less accurate for highly interbedded soil profiles with multiple liquefiable strata (e.g., Cubrinovski et al., 2019; Geyin and Maurer, 2021b; Rateria and Maurer, 2022). Of course, ML is effectively a regression technique which introduces a source of uncertainty in geospatial surrogate modeling. Geostatistical updating reduces the uncertainty introduced by ML at the locations of CPT data, yet it may introduce additional uncertainty in the near vicinity, given that the residual interpolation was constrained with a distance function and not explicitly by predictor variables. Users with proprietary geotechnical data may perform their own geostatistical updating to further constrain the predictions in their region of interest, and site-specific analysis should supersede any GLM predictions at that site. Sanger et al. (2025) provide further discussion of the limitations and uncertainties inherent to the GLMs.

*Second*, there are several sources of uncertainty in the field of ground motions input to the GLMs. This is true both for simulated earthquakes and for the spatially interpolated motions of measured events. The limitations and uncertainties of the scenario earthquakes employed in this study include source, path, and site effects. The location, geometry, and magnitude of fault rupture are uncertain, for example, as are the wave propagation and site-response models. The scenario events studied herein largely utilized the USGS ShakeMap framework, which employs a suite of ground-motion models and predicts site effects based on the estimated time-averaged shear-wave velocity ($V_{S30}$). For these reasons the predictions of liquefaction response could differ from those based on more detailed site-specific ground motion modeling. Rasanen (2025) and Rasanen et al. (2025), for example, predicted liquefaction in CSZ events at some discrete CPT sites by performing site-response analyses wherein waves were propagated through site-specific velocity structures. Users may apply the GLMs with these or other alternative $PGA_M$ values, which could lead to changes in the predicted liquefaction responses.

*Third*, there are spatial and temporal limitations in the input datasets. While the geospatial features used by the GLMs are globally available, they are often derived from models or remote sensing products with their own uncertainties and resolution constraints. Moreover, variables such as depth to groundwater and soil moisture are inherently dynamic but treated as static in this modeling framework and are inferred from the global models of Fan et al. (2013) and Hengl and Gupta (2019), respectively. Users with measured groundwater information in their area of study, or with a specific groundwater scenario of interest, may wish to interrogate the Fan et al. (2013) model in the region to understand the difference between the system being modeled (i.e., the Fan et al., 2013, groundwater surface) and the system they wish to model. Future extensions of the GLM could incorporate temporally varying inputs or real-time data streams to better reflect evolving site conditions.

Finally, impact forecasts derived from hazard simulations depend not only on accurate hazard modeling but also on reliable infrastructure and asset data, which is often sparse or inconsistently maintained. Continued development, validation, and integration of improved datasets and modeling techniques will be critical to narrowing these uncertainties and expanding the utility of the presented analyses. The impact forecasts presented in this study are demonstrative, and users interested in similar applications for their own impact forecasts should consider the coverage, source, and reliability of the infrastructure and asset data used in their analysis. For example, a utility owner wanting to assess the impact forecasts for their pipeline network will first want to compile an accurate pipeline inventory for their system. Notwithstanding these limitations, the conceptual and performance improvements exhibited by the GLMs over existing state-of-practice geospatial models make them a compelling tool for liquefaction hazard and impact simulations.

## Conclusions

This study highlights the utility of ML-based GLMs for forecasting liquefaction hazard and impacts across regional scales. The models incorporate a large suite of globally available geospatial variables and are trained to surrogate established CPT-based geotechnical models, allowing them to make physically informed predictions even in areas lacking subsurface data. Their design enables rapid deployment across many scenarios, producing probabilistic forecasts of surface disruption that are consistent with mechanistic understanding and updated where in-situ data are available. We demonstrate their use across 85 earthquake scenarios in the states of Washington and Oregon. The hazard forecasts are distributed in a GIS-ready format that supports direct application to disaster simulations, infrastructure screening, land-use planning, and emergency response. Importantly, these models reduce barriers to use by offering clear documentation and a simple computational interface, enabling diverse end-users to extract actionable insights. As such, they improve upon prior regional-scale approaches to modeling liquefaction and provide a useful tool for advancing earthquake preparedness and resilience in data-rich and data-sparse regions alike.

**Declaration of Competing Interest**

The authors declare that they have no known competing financial interests or personal relationships that could have appeared to influence the work reported in this paper.

**Acknowledgments**

This work was facilitated using high performance computing infrastructure provided by the Hyak supercomputer funded by the University of Washington's student technology fund, and through DesignSafe at the Texas Advanced Computing Center.

**Funding Sources**

The presented work is based on research supported by the United States Geological Survey (USGS) under award G23AP00017, the Cascadia Region Earthquake Science Center (CRESCENT) via National Science Foundation (NSF) award 2225286, the Cascadia Coastlines and Peoples (CoPes) Hub via NSF award 2103713, the Pacific Earthquake Engineering Research (PEER) Center under award 1185-NCTRMB, and the Pacific Northwest Transportation Consortium (PacTrans) under award 69A3552348310. However, any opinions, findings, conclusions, or recommendations expressed herein are those of the authors and may not reflect the views of USGS, NSF, CoPes, CRESCENT, PEER, or PacTrans.

**Data Availability**

The electronic results of liquefaction hazard modeling for each of the 85 scenario earthquakes are available on DesignSafe (Sanger and Maurer, 2025), including PDF maps of the results, reference data (i.e., geotiffs), and a corresponding ArcGIS Map package.

**References**

Allstadt, K.E., & Thompson, E.M. (2021). *Inventory of liquefaction features triggered by the 7 January 2020 M6.4 Puerto Rico earthquake* [Dataset]. USGS. https://doi.org/10.5066/P9HZRXI9

Allstadt, K. E., Thompson, E. M., Jibson, R. W., Wald, D. J., Hearne, M., Hunter, E. J., Fee, J., Schovanec, H., Slosky, D., & Haynie, K. L. (2022). The US Geological Survey ground failure product: Near-real-time estimates of earthquake-triggered landslides and liquefaction. *Earthquake Spectra*, *38*(1), 5-36. https://doi.org/10.1177/87552930211032685

Amatulli, G., Domisch, S., Tuanmu, M. N., Parmentier, B., Ranipeta, A., Malczyk, J., & Jetz, W. (2018). A suite of global, cross-scale topographic variables for environmental and biodiversity modeling. *Scientific data, 5*(1), 1-15. https://doi.org/10.1038/sdata.2018.40

Amatulli, G., McInerney, D., Sethi, T., Strobl, P., & Domisch, S. (2020). Geomorpho90m, empirical evaluation and accuracy assessment of global high-resolution geomorphometric layers. *Scientific Data, 7*(162). https://doi.org/10.1038/s41597-020-0479-6

Anselin, L. (1995). The Local Indicators of Spatial Association – LISA. *Geographical Analysis*, 27, 93-115. https://doi.org/10.1111/j.1538-4632.1995.tb00338.x

Asadi, A., Sanon, C., Cakir, E., Zhan, W., Shirzadi, H., Baise, L. G., Cetin, K. O., & Moaveni, B. (2024). Geospatial liquefaction modeling of the 2023 Türkiye earthquake sequence by an ensemble of global, continental, regional, and event-specific models. *Seismological Research Letters*, *95*(2A), 697-719. https://doi.org/10.1785/0220230287

Azul, K. M., Orense, R., & Wotherspoon, L. (2024). Investigation of representative geotechnical data for the development of a hybrid geotechnical-geospatial liquefaction assessment model. *Japanese Geotechnical Society Special Publication*, *10*(17), 585-590. https://doi.org/10.3208/jgssp.v10.OS-6-07

Bozzoni, F., Bonì, R., Conca, D., Meisina, C., Lai, C. G., & Zuccolo, E. (2021). A geospatial approach for mapping the earthquake-induced liquefaction risk at the European scale. *Geosciences*, *11*(1), 32. https://doi.org/10.3390/geosciences11010032

Bray, J. D., Sancio, R. B., Kammerer, A. M., Merry, S., Rodriguez-Marek, A., Khazai, B., Chang, S., Bastani, A., Collins, B., Hausler, E., Dreger, D., Perkins, W. J., & Nykamp, M. (2001). Some observations of geotechnical aspects of the February 28, 2001 Nisqually earthquake in Olympia, South Seattle, and Tacoma, Washington. *Report sponsored by NSF, PEER Center, UCB, University of Arizona, Washington State University, Shannon and Wilson Inc., and Leighton and Associates*.

Breiman, L. (1996), Bagging predictor. *Machine Learning, 24*(2), 123-140. https://doi.org/10.1007/BF00058655

Breiman, L. (1999). Pasting small votes for classification in large databases and on-line. *Machine Learning* 36, 85–103 https://doi.org/10.1023/A:1007563306331

British Red Cross. (2024). *Türkiye (Turkey) and Syria earthquake 2023: a year on*. British Red Cross. Retrieved June 10, 2025, from https://www.redcross.org.uk/stories/disasters-and-emergencies/world/turkey-syria-earthquake

Brown, J., Ferrians, O., Heginbottom, J. A., & Melnikov, E. (2002). *Circum-Arctic map of permafrost and ground-ice conditions, Version 2* [Dataset]. NASA National Snow and Ice Data Center Distributed Active Archive Center. https://doi.org/10.7265/skbgkf16

Bullock, Z., Zimmaro, P., Lavrentiadis, G., Wang, P., Ojomo, O., Asimaki, D., Rathje, E. M., & Stewart, J. P. (2023). A latent Gaussian process model for the spatial distribution of liquefaction manifestation. *Earthquake Spectra*, *39*(2), 1189-1213. https://doi.org/10.1177/87552930231163894

Cetin, K. O., Soylemez, B., Guzel, H., & Cakir, E. (2024). Soil liquefaction sites following the February 6, 2023, Kahramanmaraş-Türkiye earthquake sequence. *Bulletin of Earthquake Engineering*, 1-24. https://doi.org/10.1007/s10518-024-01875-3

Cohen, J. (1988). *Statistical power analysis for the behavioral sciences*. Hillsdale, NJ: Lawrence Erlbaum.

Columbia University Center For International Earth Science Information Network (CIESIN). (2017). *Gridded Population of the World, Version 4 (GPWv4): Population Density, Revision 11* (Version 4.11) [Dataset]. Palisades, NY: Socioeconomic Data and Applications Center (SEDAC). https://doi.org/10.7927/H49C6VHW

Cubrinovski, M., Rhodes, A., Ntritsos, N., & Van Ballegooy, S. (2019). System response of liquefiable deposits. *Soil Dynamics and Earthquake Engineering*, *124*, 212-229. https://doi.org/10.1016/j.soildyn.2018.05.013

Fan, Y., Li, H., & Miguez-Macho, G. (2013, updated 2019). Global patterns of groundwater table depth. *Science, 339*, 940-943. https://www.science.org/doi/10.1126/science.1229881

Fawcett, T. (2006). An introduction to ROC analysis. *Pattern recognition letters, 27*(8), 861-874. https://doi.org/10.1016/j.patrec.2005.10.010

Frankel, A., Wirth, E., Marafi, N., Vidale, J., & Stephenson, W. (2018). Broadband synthetic seismograms for magnitude 9 earthquakes on the Cascadia megathrust based on 3D simulations and stochastic synthetics, part 1: Methodology and overall results. *Bulletin of the Seismological Society of America, 108*(5A), 2347-2369. https://doi.org/10.1785/0120180034

Geyin, M., & Maurer, B. W. (2020). Fragility functions for liquefaction-induced ground failure. *Journal of Geotechnical and Geoenvironmental Engineering*, *146*(12), 04020142. https://doi.org/10.1061/(ASCE)GT.1943-5606.0002416

Geyin, M., & Maurer, B. W. (2021a). *CPT-Based Liquefaction Case Histories from Global Earthquakes: A Digital Dataset (Version 1)* [Dataset]. DesignSafe-CI. https://doi.org/10.17603/ds2-wftt-mv37

Geyin, M., & Maurer, B. W. (2021b). Evaluation of a cone penetration test thin-layer correction procedure in the context of global liquefaction model performance. *Engineering Geology*, *291*, 106221. https://doi.org/10.1016/j.enggeo.2021.106221

Geyin, M., Baird, A.J. & Maurer, B.W. (2020). Field assessment of liquefaction prediction models based on geotechnical vs. geospatial data, with lessons for each. *Earthquake Spectra, 36*(3), 1386–1411. https://doi.org/10.1177/875529301989995

Geyin, M., Maurer, B. W., & Christofferson, K. (2022). An AI driven, mechanistically grounded geospatial liquefaction model for rapid response and scenario planning. *Soil Dynamics and Earthquake Engineering*, *159*, 107348. https://doi.org/10.1016/j.soildyn.2022.107348

Heath, D., Wald, D. J., Worden, C. B., Thompson, E. M., & Smoczyk, G. (2020). A global hybrid Vs30 map with a topographic-slope-based default and regional map insets. *Earthquake Spectra, 36*(3), 1570-1584.

Hengl, T. (2018a). *Soil bulk density (fine earth) 10 x kg / m-cubic at 6 standard depths (0, 10, 30, 60, 100 and 200 cm) at 250 m resolution (v0.2)* [Dataset]. Zenodo. https://doi.org/10.5281/zenodo.2525665

Hengl, T. (2018b). *Clay content in % (kg / kg) at 6 standard depths (0, 10, 30, 60, 100 and 200 cm) at 250 m resolution (v0.2)* [Dataset]. Zenodo. https://doi.org/10.5281/zenodo.2525663

Hengl, T. (2018c). *Sand content in % (kg / kg) at 6 standard depths (0, 10, 30, 60, 100 and 200 cm) at 250 m resolution (v0.2)* [Dataset]. Zenodo. https://doi.org/10.5281/zenodo.2525662

Hengl, T. (2018d). *Silt content in % (kg / kg) at 6 standard depths (0, 10, 30, 60, 100 and 200 cm) at 250 m resolution (v0.2)* [Dataset]. Zenodo. https://doi.org/10.5281/zenodo.2525676

Hengl, T., & Gupta, S. (2019). *Soil water content (volumetric %) for 33kPa and 1500kPa suctions predicted at 6 standard depths (0, 10, 30, 60, 100 and 200 cm) at 250 m resolution (v0.1)* [Dataset]. Zenodo. https://doi.org/10.5281/zenodo.2784001

Hengl, T., & Nauman, T. (2018). Predicted USDA soil great groups at 250 m (probabilities) (v0.2) [Data set]. Zenodo. https://doi.org/10.5281/zenodo.352806

Hengl, T., Heuvelink, G. B., & Rossiter, D. G. (2007). About regression-kriging: From equations to case studies. *Computers & geosciences*, *33*(10), 1301-1315. https://doi.org/10.1016/j.cageo.2007.05.001

Hu, J., & Liu, H. (2019). Bayesian network models for probabilistic evaluation of earthquake-induced liquefaction based on CPT and Vs databases. *Engineering Geology, 254*, 76-88. https://doi.org/10.1016/j.enggeo.2019.04.003

Idriss, I. M., & Boulanger, R. W. (2008). *Soil liquefaction during earthquakes*. Earthquake Engineering Research Institute.

Iwasaki, T. (1978). A practical method for assessing soil liquefaction potential based on case studies at various sites in Japan. In *Proc. of 2nd Int. National Conf. on Microzonation, 1978* (Vol. 2, pp. 885-896).

Jaiswal, K. S., Rozelle, J., Tong, M., Sheehan, A., McNabb, S., Kelly, M., Zuzak, C., Bausch, D., & Sims, J. (2023). Hazus Estimated Annualized Earthquake Losses for the United States: FEMA P-366, 2023. *USGS Report*, 8.

Jena, R., Pradhan, B., Almazroui, M., Assiri, M., & Park, H. J. (2023). Earthquake-induced liquefaction hazard mapping at national-scale in Australia using deep learning techniques. *Geoscience Frontiers*, *14*(1), 101460. https://doi.org/10.1016/j.gsf.2022.101460

Kim, H. S. (2023). Geospatial data-driven assessment of earthquake-induced liquefaction impact mapping using classifier and cluster ensembles. *Applied Soft Computing*, *140*, 110266. https://doi.org/10.1016/j.asoc.2023.110266

Lehner, B., & Grill G. (2013). Global river hydrography and network routing: baseline data and new approaches to study the world's large river systems. *Hydrological Processes, 27*(15): 2171–2186. https://doi.org/10.1002/hyp.9740

Lewis, S. (2009). *Hydrologic Sub-basins of Greenland (NSIDC-0371, Version 1)* [Dataset]. NASA National Snow and Ice Data Center Distributed Active Archive Center. https://doi.org/10.5067/DT9T7DPD7HBI

Maurer, B. W., & Sanger, M. D. (2023). Why "AI" models for predicting soil liquefaction have been ignored, plus some that shouldn't be. *Earthquake Spectra*, *39*(3), 1883-1910. https://doi.org/10.1177/87552930231173711

Maurer, B. W., Geyin, M., & Van Ballegooy, S. (2025). A pragmatic framework for cost-benefit analysis of liquefaction mitigation applicable to lightweight residential structures on shallow foundations. *Journal of Geotechnical and Geoenvironmental Engineering*, *151*(6), 04025045. https://doi.org/10.1061/JGGEFK.GTENG-13142

Maurer, B. W., Green, R. A., & Taylor, O. D. S. (2015). Moving towards an improved index for assessing liquefaction hazard: Lessons from historical data. *Soils and foundations*, *55*(4), 778-787. https://doi.org/10.1016/j.sandf.2015.06.010

Messager, M. L., Lehner, B., Grill, G., Nedeva, I., & Schmitt, O. (2016). Estimating the volume and age of water stored in global lakes using a geo-statistical approach. *Nature Communications, 7*, 13603. https://doi.org/10.1038/ncomms13603

Ministry for Culture and Heritage. (2023). *Christchurch earthquake kills 185: 22 February 2011*. NZHistory. Retrieved June 10, 2025, from https://nzhistory.govt.nz/page/christchurch-earthquake-kills-185

NASA. (2020) *Distance to nearest coastline.* [Dataset] NASA Ocean Biol Process Group (OBPG) 2. https://oceancolor.gsfc.nasa.gov/resources/docs/distfromcoast/

New Zealand Earthquake Commission (EQC). (2016). *New Zealand Geotechnical Database (NZGD)* [Dataset]. https://www.nzgd.org.nz/

Nobre, A. D., Cuartas, L. A., Hodnett, M., Rennó, C. D., Rodrigues, G., Silveira, A., & Saleska, S. (2011). Height Above the Nearest Drainage–a hydrologically relevant new terrain model. *Journal of Hydrology, 404*(1-2), 13-29. https://doi.org/10.1016/j.jhydrol.2011.03.051

Oregon Department of Transportation (ODOT). (2025). *TransGIS* [Dataset]. Geographic Information Services Unit. https://gis.odot.state.or.us/transgis/

Parker, M., & Steenkamp, D. (2012). The economic impact of the Canterbury earthquakes. *Reserve Bank of New Zealand Bulletin, 75*(3), 13-25.

Petersen, M.D., Moschetti, M. P., Powers, P. M., Mueller, C. S., Haller, K. M., Frankel, A. D., Zeng, Y., Rezaeian, S., Harmsen, S. C., Boyd, O. S., Field, N., Chen, R., Rukstales, K. S., Luco, N., Wheeler, R. L., Williams, R. A., & Olsen, A. H. (2014). Documentation for the 2014 Update of the United States National Seismic Hazard Maps: U.S. Geological Survey Open-File Report 2014–1091. http://dx.doi.org/10.3133/ofr20141091

Rasanen, R. A., Marafi, N. A., & Maurer, B. W. (2021). Compilation and forecasting of paleoliquefaction evidence for the strength of ground motions in the US Pacific Northwest. *Engineering Geology*, *292*, 106253. https://doi.org/10.1016/j.enggeo.2021.106253

Rasanen, R. A., Geyin, M., & Maurer, B.W. (2022). *Select CPT-Based Liquefaction Case Histories from the 2001 Nisqually, Washington, Earthquake* [Dataset]. DesignSafe-CI. https://doi.org/10.17603/ds2-nsf8-7944

Rasanen, R. A., Geyin, M., & Maurer, B.W. (2023). Select liquefaction case histories from the 2001 Nisqually, Washington earthquake: A digital dataset and assessment of model performance. *Earthquake Spectra*, *39*(3): 1534-1557. https://doi.org/10.1177/87552930231174244

Rasanen, R. A., Geyin, M., Sanger, M. D., & Maurer, B. W. (2024). *A database of cone penetration tests from the Cascadia Subduction Zone* [Dataset]. DesignSafe-CI. https://doi.org/10.17603/ds2-snvw-jv27

Rasanen, R. A., Grant, A. R., Makdisi, A. J., Maurer, B. W., & Wirth, E. A. (2025). Implications of physics-based M9 ground motions on liquefaction-induced damage in the Cascadia Subduction Zone: Looking forward and backward. *Earthquake Spectra*, *41*(2), 999-1028. https://doi.org/10.1177/87552930251316819

Rashidian, V., & Baise, L. G. (2020). Regional efficacy of a global geospatial liquefaction model. *Engineering Geology*, *272*, 105644. https://doi.org/10.1016/j.enggeo.2020.105644

Rateria, G., & Maurer, B. W. (2022). Evaluation and updating of Ishihara's (1985) model for liquefaction surface expression, with insights from machine and deep learning. *Soils and Foundations*, *62*(3), 101131. https://doi.org/10.1016/j.sandf.2022.101131.

Rathje, E., Dawson, C. Padgett, J.E., Pinelli, J.-P., Stanzione, D., Adair, A., Arduino, P., Brandenberg, S.J., Cockerill, T., Dey, C., Esteva, M., Haan, Jr., F.L., Hanlon, M., Kareem, A., Lowes, L., Mock, S., & Mosqueda, G. (2017). DesignSafe: A new cyberinfrastructure for natural hazards engineering. *Natural Hazards Review, 18*(3). https://doi.org/10.1061/(ASCE)NH.1527-6996.0000246

Regione Emilia-Romagna. (2024). *Banca dati prove geognostiche - Prove penetrometriche numeriche (punti)* [Dataset]. Geoportale. https://geoportale.regione.emilia-romagna.it/catalogo/dati-cartografici/informazioni-geoscientifiche/geologia/banca-dati-geognostica/layer-3

RGI Consortium. (2023). *Randolph Glacier Inventory - A Dataset of Global Glacier Outlines (NSIDC-0770, Version 7).* [Dataset]. National Snow and Ice Data Center. https://doi.org/10.5067/F6JMOVY5NAVZ

Sanger, M. D., Geyin, M., Shin, A., & Maurer, B. W. (2024a). *A database of cone penetration tests from North America* [Dataset]. DesignSafe-CI. https://doi.org/10.17603/ds2-gqjm-t836

Sanger, M. D., Geyin, M., Maurer, B. W. (2024b). *Mechanics-informed machine learning for geospatial modeling of soil liquefaction: Global model map products for LPI, LPIish, and LSN* [Dataset]. DesignSafe-CI. https://doi.org/10.17603/ds2-c0z7-hc12

Sanger, M. D., Geyin, M., Maurer, B. W. (2024c). *Mechanics-informed machine learning for geospatial modeling of soil liquefaction: Example model implementation in Jupyter Notebook and Matlab* [Dataset]. DesignSafe-CI. https://doi.org/10.17603/ds2-sp3e-dp21

Sanger, M.D., Geyin, M., & Maurer, B.W. (2025). Mechanics-informed machine learning for geospatial modeling of soil liquefaction: global and national surrogate models for simulation and near-real-time response. *ASCE Journal of Geotechnical and Geoenvironmental Engineering, 151*(11): 04025126. https://doi.org/10.1061/JGGEFK.GTENG-13737

Sanger, M. D., & Maurer, B. W. (2025). *Pacific Northwest liquefaction hazard simulations* [Dataset]. DesignSafe-CI. https://doi.org/10.17603/ds2-1ryc-0w40

Shangguan, W., Hengl, T., Mendes de Jesus, J., Yuan, H., & Dai, Y. (2017). Mapping the global depth to bedrock for land surface modeling. *Journal of Advances in Modeling Earth Systems, 9*(1), 65-88. https://doi.org/10.1002/2016MS000686

Smirnov, N. V. (1939). On the estimation of the discrepancy between empirical curves of distribution for two independent samples. *Bull. Math. Univ. Moscou, 2*(2), 3-14.

Steinbach, M. (2000). A comparison of document clustering techniques. In *KDD Workshop on Text Mining, 2000.*

Taftsoglou, M., Valkaniotis, S., Papathanassiou, G., & Karantanellis, E. (2023). Satellite imagery for rapid detection of liquefaction surface manifestations: The case study of Türkiye–Syria 2023 earthquakes. *Remote Sensing, 15*, 4190. https://doi.org/10.3390/rs15174190

Todorovic, L., & Silva V. (2022). A liquefaction occurrence model for regional analysis. *Soil Dynamics and Earthquake Engineering, 161*(1), 107430. https://doi.org/10.1016/j.soildyn.2022.107430

Toprak, S., Nacaroglu, E., Van Ballegooy, S., Koc, A. C., Jacka, M., Manav, Y., Torvelainen, E., & O'Rourke, T. D. (2019). Segmented pipeline damage predictions using liquefaction vulnerability parameters. *Soil Dynamics and Earthquake Engineering*, *125*, 105758. https://doi.org/10.1016/j.soildyn.2019.105758

Ulmer, K. J., Zimmaro, P., Brandenberg, S. J., Stewart, J. P., Hudson, K. S., Stuedlein, A. W., Jana, A., Dadashiserej, A., Kramer, S. L., Cetin, K. O., Can, G., Ilgac, M., Franke, K. W., Moss, R. E. S., Bartlett, S. F., Hosseinali, M., Dacayanan, H., Kwak, D. Y., Stamatakos, J., Mukherjee, J., Salman, U., Ybarra, S., & Weaver, T. (2023). *Next-Generation Liquefaction* [Database, Version 2]. Next-Generation Liquefaction Consortium. https://doi.org/10.21222/C23P70

Ulmer, K. J., Hudson, K. S., Brandenberg, S. J., Zimmaro, P., Pretell, R., Carlton, B., Kramer, S. L., & Stewart, J. P. (2024). Next Generation Liquefaction models for susceptibility, triggering, and manifestation (Rev. 1; Research Information Letter RIL2024-13). U.S. Nuclear Regulatory Commission, Office of Nuclear Regulatory Research. https://www.nrc.gov/docs/ML2435/ML24353A158.pdf

Upadhyaya, S., Green, R. A., Maurer, B. W., Rodriguez-Marek, A., & van Ballegooy, S. (2022). Limitations of surface liquefaction manifestation severity index models used in conjunction with simplified stress-based triggering models. *ASCE Journal of Geotechnical and Geoenvironmental Engineering, 148*(3). https://doi.org/10.1061/(ASCE)GT.1943-5606.0002725

U.S. Army Corps of Engineers (USACE). (2024). *National Inventory of Dams* [Dataset]. https://nid.sec.usace.army.mil/#/

U.S. Department of Transportation (USDOT), Federal Highway Administration (FHWA); Bureau of Transportation Statistics (BTS). (2020). National Bridge Inventory 2008-Present [Dataset]. https://doi.org/10.21949/15

USGS. (2019). *Cone Penetration Testing (CPT)* [Dataset]. Earthquake Hazards Program. https://www.usgs.gov/programs/earthquake-hazards/science/cone-penetration-testing-cpt#overview

USGS. (n.d.-a). *BSSC2014 (Scenario Catalog): Building Seismic Safety Council 2014 Event Set.* U.S. Geological Survey. https://earthquake.usgs.gov/scenarios/catalog/bssc2014

USGS. (n.d.-b). *CSZM9 (Scenario Catalog:M9 Cascadia Earthquake Scenarios.* U.S. Geological Survey. https://earthquake.usgs.gov/scenarios/catalog/cszm9/

USGS. (n.d.-c). *WA22 (Scenario Catalog):ShakeMap Scenario Catalog for Selected Quaternary Active Faults in Washington State.* U.S. Geological Survey. https://earthquake.usgs.gov/scenarios/catalog/wa22

Van Ballegooy, S., Malan, P., Lacrosse, V., Jacka, M. E., Cubrinovski, M., Bray, J. D., O'Rourke, T. D., Crawford, S.A., & Cowan, H. (2014). Assessment of liquefaction-induced land damage for residential Christchurch. *Earthquake Spectra*, *30*(1), 31-55. https://doi.org/10.1193/031813EQS070M

Wang, J., & Hu, J. (2025). A GLPI framework for gravelly soil liquefaction hazard assessment based on fuzzy mathematics. *Engineering Geology, 353*, 108134. https://doi.org/10.1016/j.enggeo.2025.108134

Wirth, E. A., Grant, A., Marafi, N. A., & Frankel, A. D. (2021). Ensemble ShakeMaps for magnitude 9 earthquakes on the Cascadia subduction zone, *Seismological Research Letters*, *92*(1), 199-211. https://doi.org/10.1785/0220200240

Zhu, J., Baise, L. G., & Thompson, E. M. (2017). An updated geospatial liquefaction model for global application. *Bulletin of the Seismological Society of America*, *107*(3): 1365-1385. https://doi.org/10.1785/0120160198

Zimmaro, P., Nweke, C. C., Hernandez, J. L., Hudson, K. S., Hudson, M. B., Ahdi, S. K., Boggs, M. L., Davis, C. A., Goulet, C. A., Brandenburg, S. J., Hudnut., K. W., & Stewart, J. P. (2020). Liquefaction and related ground failure from July 2019 Ridgecrest earthquake sequence. *Bulletin of the Seismological Society of America*, *110*(4), 1549-1566. https://doi.org/10.1785/0120200025